\documentclass[twocolumn]{aastex6}
\usepackage{amsmath}
\usepackage{amssymb}
\usepackage{graphicx}
\usepackage{microtype}
\usepackage{breakcites}
\usepackage{hyperref}

\shortauthors{Roth et al.}
\shorttitle{x-ray through optical emission from TDEs}

\begin{document}

\title{The X-ray through Optical Fluxes and Line Strengths of Tidal Disruption Events}

\author{Nathaniel Roth\altaffilmark{1}}
\author{Daniel Kasen\altaffilmark{1,2,3}}
\author{James Guillochon\altaffilmark{4,5}}
\author{Enrico Ramirez-Ruiz\altaffilmark{6}}
\email{nathaniel.roth@berkeley.edu}
\altaffiltext{1}{Physics Department, University of California, Berkeley, CA
94720, USA}
\altaffiltext{2}{Astronomy Department and Theoretical Astrophysics Center,
University of California, Berkeley, CA 94720, USA}
\altaffiltext{3}{Nuclear Science Division, Lawrence Berkeley National Laboratory,
Berkeley, CA 94720, USA}
\altaffiltext{4}{Harvard-Smithsonian Center for Astrophysics, The Institute for Theory
and Computation, 60 Garden Street, Cambridge, MA 02138, USA}
\altaffiltext{5}{Einstein Fellow}
\altaffiltext{6}{Department of Astronomy and Astrophysics, University of California Santa Cruz, 1156 High Street, Santa Cruz, CA 95060, USA}

\begin{abstract} 
We study the emission from tidal disruption events (TDEs) produced as radiation from black hole accretion propagates through an extended, optically thick envelope formed from stellar debris. We analytically describe key physics controlling spectrum formation, and present detailed radiative transfer calculations that model the spectral energy distribution (SED) and optical line strengths of TDEs near peak brightness. The steady-state transfer is coupled to a solver for the excitation and ionization states of hydrogen, helium and oxygen (as a representative metal), without assuming local thermodynamic equilibrium. Our calculations show how an extended envelope can reprocess a fraction of soft x-rays and produce the observed optical fluxes of order $10^{43}~{\rm erg~s^{-1}}$, with an optical/UV continuum that is not described by a single blackbody. Variations in the mass or size of the envelope may help explain how the optical flux changes over time with roughly constant color. For high enough accretion luminosities, x-rays can escape to be observed simultaneously with the optical flux. Due to optical depth effects, hydrogen Balmer line emission is often strongly suppressed relative to helium line emission (with HeII-to-H line ratios of at least 5:1 in some cases) even in the disruption of a solar-composition star. We discuss the implications of our results to understanding the type of stars destroyed in TDEs and the physical processes responsible for producing the observed flares.
\end{abstract}


\newcommand{\alphaR}{\ensuremath{\alpha_{\rm r}}}

\newcommand{\rs}{\ensuremath{r_{\rm s}}}

\newcommand{\Ha}{\ensuremath{{\rm H}\alpha}}

\newcommand{\rin}{\ensuremath{r_{\rm in}}}
\newcommand{\rins}{\ensuremath{r_{\rm i,14}}}
\newcommand{\rout}{\ensuremath{r_{\rm out}}}
\newcommand{\routs}{\ensuremath{r_{\rm o,15}}}
\newcommand{\rhoin}{\ensuremath{\rho_{\rm in}}}
\newcommand{\nin}{\ensuremath{n_{\rm in}}}

\newcommand{\Menv}{\ensuremath{M_{\rm env}}}
\newcommand{\Menvs}{\ensuremath{M_{\rm e,0.5}}}
\newcommand{\msun}{\ensuremath{M_\odot}}
\newcommand{\taues}{\ensuremath{\tau_{\rm es}}}
\newcommand{\Ep}{\ensuremath{\bar{E}_{\rm s}}}
\newcommand{\Eps}{\ensuremath{\bar{E}_{\rm s,50}}}

\newcommand{\Erad}{\ensuremath{E_{\rm rad}}}
\newcommand{\Trad}{\ensuremath{T_{\rm rad}}}
\newcommand{\Tgas}{\ensuremath{T_{\rm gas}}}
\newcommand{\Ts}{\ensuremath{T_{\rm s}}}
\newcommand{\Ls}{\ensuremath{L}}
\newcommand{\Lr}{\ensuremath{L_{\rm r}}}
\newcommand{\Lion}{\ensuremath{L_{\rm ion}}}

\newcommand{\Tr}{\ensuremath{T_{\rm r}}}

\newcommand{\Tss}{\ensuremath{T_{\rm s,5}}}

\newcommand{\Teff}{\ensuremath{T_{\rm eff}}}
\newcommand{\vopt}{\ensuremath{\nu_{\rm opt}}}
\newcommand{\nuion}{\ensuremath{\nu_{\rm ion}}}

\newcommand{\gcc}{\ensuremath{{\rm g~cm^{-3}}}}
\newcommand{\csg}{\ensuremath{{\rm cm^{2}~g^{-1}}}}
\newcommand{\kms}{\ensuremath{{\rm km~s^{-1}}}}
\newcommand{\ergss}{\ensuremath{{\rm erg~s^{-1}}}}

\section{Introduction}
\label{sec:Introduction}

If a star passes deep enough within the gravitational potential of a black hole (BH), tidal forces can exceed self-gravity, ripping the star apart in a tidal disruption event (TDE).   Subsequent shocks occurring in colliding stellar debris streams \citep[e.g.][]{Kochanek1994,Guillochon2015-1}, and/or the eventual accretion of this gas onto the BH may produce
 a luminous thermal flare at x-ray through optical wavelengths.  A relativistic jet generated by the accreting BH may also produce non-thermal gamma-ray, x-ray and radio emission.

Observational candidates for TDEs are rapidly accumulating. A number of flares from galactic centers have been discovered in x-rays \citep{Komossa1999,Donley2002,Komossa2004,Halpern2004,Esquej2007,Cappelluti2009,Maksym2010,Saxton2012,Hryniewicz2016,Lin2015,Komossa2015}.
The peak luminosity is high, $\gtrsim 10^{44}$ erg s$^{-1}$, and the spectral energy distribution (SED) peaks at soft x-ray energies $\lesssim 0.1$ keV.
After peak, the luminosity fades as a power-law in time similar to $L \propto t^{-5/3}$, a dependence predicted for the fallback of disrupted stellar debris \citep{Rees1988,Phinney1989,Evans1989,Lodato2009,Guillochon2013-1}. 

There have also been TDE candidates found in the ultraviolet (UV) \citep{Gezari2006,Gezari2009}, and in the optical in SDSS \citep{van-Velzen2011,van-Velzen2014},  Pan-STARRS1 \citep{Gezari2012,Chornock2014}, ASASSN \citep{Holoien2014,Holoien2016-1,Holoien2016-2}, PTF \citep{Cenko2012-1,Arcavi2014}, and ROTSE \citep{Vinko2015}. These events typically rise to a peak (observer frame) R-band luminosity of $\sim 2 \times 10^{43}~{\rm ergs~s^{-1}}$ on a timescale of $\sim$ months \citep{Arcavi2014}, with a late time fall consistent with $t^{-5/3}$.  Intriguingly, PTF10iya, Swift J2058.4,  ASASSN-14li, and ASASSN-15oi have been simultaneously observed in both optical and x-ray \citep{Cenko2012-1,Cenko2012-2,Holoien2016-1,Miller2015-2,Cenko2016,Holoien2016-2}.  

  A few long-lived ($\sim 10^7$~s) transients have also been observed
  at hard x-ray and gamma-ray energies \citep{Bloom2011,Cenko2012-2,Brown2015}. Occasionally, associated optical
emission is also detected.  
These events have been interpreted as due to a relativistic jet generated via BH accretion; 
such ``jetted'' TDEs appear 
to be observationally rare compared to the soft x-ray and UV/optical flares  \citep{van-Velzen2014,Arcavi2014}, in line with theoretical expectations \citep{De-Colle2012}.

While many aspects of the TDE candidates remain poorly understood, the nature of the optical/UV emission is perhaps the most puzzling. Two fundamental questions  await full explanation: 1) Why is the  observed optical flux in the UV/optical transients orders of magnitudes higher than that predicted by a standard BH accretion disk,  and with a blue 
color that remains roughly constant over time? 2) Why do the optical spectra show strong lines of helium, but little or no hydrogen line emission \citep{Gezari2012, Arcavi2014}?

The first puzzle stems from the fact that the tidal disruption radius, $R_{\rm td} = (M_{\rm bh}/M_\star)^{1/3} R_\star$, is  $\approx 10^{13}$~cm for the
disruption of a solar-like star (mass $M_\star = M_\odot$,
radius $R_\odot$) by a  BH  of mass $M_{\rm bh} \sim 10^6~M_\odot$.   Thermal
emission from this radius should be in the soft x-ray (temperatures $\gtrsim10^5$~K)
with low optical luminosity ($\lesssim 10^{42}~{\rm ergs~s^{-1}})$.  The problem has been addressed by postulating the presence of gas at large radii ($\sim 100$ times $R_{\rm td}$) that  absorbs (or advects)  radiation and re-emits it at lower temperatures of a few times $10^4$~K.
This reprocessing region may be due to the formation of a hydrostatic (or quasi-static) envelope around the BH \citep{Loeb1997,Guillochon2014,Coughlin2014-1}, or a super-Eddington mass outflow \citep{Strubbe2009,Lodato2011,Metzger2015,Vinko2015,Miller2015-1}, or the circularization of material at radii much greater $R_{\rm td}$ \citep{Shiokawa2015,Piran2015,Hayasaki2015,Bonnerot2016-1,Guillochon2015-1,Dai2015}.  As of yet, however, no detailed radiative transfer 
calculations have determined how, or if, TDE radiation can be so reprocessed, and if so how the emergent  optical through x-ray emission depends on the gas properties.  These are key questions we hope to address here. 

The solution to the second puzzle  -- the low hydrogen to helium emission line ratios -- continues to be debated. Early theoretical modeling of the \Ha\ emission by \citet{Bogdanovic2004} raised the possibility that this line emission might be destroyed due to the high optical depth to \Ha\ photons in a reprocessing region surrounding the inflowing stellar debris. \citet{Gezari2012} argued that the absence of H lines in PS1-10jh implies the disruption of a He-rich stellar core. The simulations of \citet{Macleod2012}, however, show that it is difficult to remove the H envelope of a red giant and disrupt only the He core.  Noting that He star disruptions would be exceedingly rare, \citet{Guillochon2014} instead argued for the disruption of a normal main sequence star but with
the hydrogen line emission suppressed by photoionization effects \citep[e.g.,][]{Korista2004}. \citet{Gaskell2014} performed new calculations using the photoionization code CLOUDY \citep{Ferland1998,Ferland2013} and showed that, if the lines are optically thick, radiation transport effects  may reduce the hydrogen emission line strength.  However, a separate CLOUDY parameter study performed by \citet{Strubbe2015} disputed this interpretation, concluding that, for the conditions relevant to TDEs, hydrogen lines would not be suppressed enough to be consistent with the observations of PS1-10jh.

All previous studies of TDE  line emission, however, have been subject to limitations of photoionization codes like CLOUDY, which assume that the gas is optically thin in the continuum (but may allow for optical depth effects for lines).  Here we show that TDE envelopes are likely optically thick and highly scattering dominated, and that can profoundly change the nature of line  and continuum formation.   We first use analytic arguments to delineate  the
 physical conditions and key radiative processes in TDE envelopes, and then
carry out the first  non-local-thermodynamic equilibrium (non-LTE) 
Boltzmann radiative transfer calculations for an idealized TDE model with a spherically, optically thick reprocessing region. 

\begin{figure*}
\includegraphics[width=\textwidth]{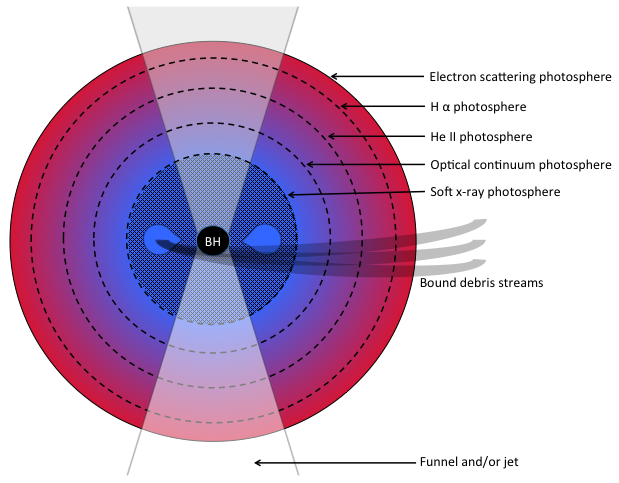}
\caption{A schematic (not entirely to-scale) representation of a quasi-static TDE
envelope, marking the \emph{effective} photospheres at different wavelengths
(beyond the effective photosphere, most photons will electron scatter their way out of the envelope without being re-absorbed). The soft x-ray photosphere lies deepest in,
perhaps near the tidal disruption radius.  The optical continuum photosphere lies farther out, followed by the He II~$\lambda4686$ photosphere. The \Ha\ photosphere lies farthest out,  near to the Thomson photosphere. The larger visible volume of  helium line emission results in a high He to H emission line ratio  in 
the model TDE spectra.  Although our calculations assume spherical symmetry, we have suggested the presence of possible viewing angle effects owing to the presence of a funnel or jet in the polar directions.}
\label{fig:schematic}
\end{figure*}

In contrast to  optically thin photoionization models, spectrum formation in optically thick TDE envelopes more closely resembles the situation in stellar atmospheres, where the emission at a given wavelength depends on the source function at the associated thermalization depth. 
The thermalization depth is the radius at which emitted photons can scatter to the surface without being reabsorbed, and corresponds
to a radial optical depth (integrated inward) of approximately $\sqrt{ \tau_{\rm abs} \taues}$ where $\tau_{\rm abs}$  and \taues\ are the optical depths to
absorptive and electron scattering processes, respectively. 
The thermalization depth varies with wavelength, such that lines and the continuum form at different layers (see Figure~\ref{fig:schematic}).
The resulting spectrum will generally not be described by a single blackbody.

Our calculations define the conditions under which a TDE envelope may be ''effectively'' optically thick  (i.e., $\sqrt{ \tau_{\rm abs} \taues} \gtrsim 1)$ at soft x-ray wavelengths, and so absorb and reprocess a fraction of the accretion luminosity into optical  continuum and line emission.
 We describe how the reprocessing efficiency depends on  the mass, size, and ionization state of the envelope.  Variations in these quantities may help explain the observed optical flux evolution and the occasional simultaneous appearance of x-rays.  We show that escaping line photons are produced only in the outer layers of the envelope, such that the H to He line ratio is suppressed and will vary with the envelope extent.   
 Indeed,  some optical TDE candidates do show detectable, but varying, H$\alpha$ emission \citep{Arcavi2014,Holoien2016-1}.   
 
In Section~\ref{sec:analytics}, we describe the assumptions of the model setup, and make analytic estimates of the degree of reprocessed luminosity. In Section~\ref{sec:Results} we present our numerical results from our  radiative transport calculations. In Section~\ref{sec:ObservationDiscussion} we discuss implications for observations and models of TDEs, and in Section~\ref{sec:Conclusions} we conclude with a summary of the most important take away points.

\section{Analytic Considerations}
\label{sec:analytics}

An accurate determination of the optical/UV luminosity from TDE envelopes requires a detailed non-LTE radiative transfer calculation, which we provide
in Section~\ref{sec:Results}.  However, some insight into the physics of radiation reprocessing, and how it depends on envelope parameters,
can be gained by approximate analytic arguments. 

We consider a 1D configuration where radiation is emitted by a spherical source at radius \rin\ defining the layer at which most of the  radiative luminosity is  generated.  Choosing \rin\ comparable to the original stellar pericenter passage distance (typically $\sim 10^{13}$~cm for a $10^6$-$10^7$ \msun\ BH) would approximate the luminosity arising from a viscous accretion disk formed near the disruption radius.  
Alternatively, the luminosity could be generated by the circularization of fallback material at a larger radius, $\rin \sim 10^{14}$~cm \citep[e.g.][]{Piran2015,Dai2015}.

We assume that the inner boundary radiates blackbody radiation at a source temperature \Ts.  Our implicit assumption is that
the conditions interior to \rin\ are sufficient to thermalize the radiation, e.g., via Comptonization or adiabatic expansion.
For luminosities of order $10^{44}-10^{45}~\ergss$, $\Ts \sim 10^5- 10^6$~K and source photons are emitted primarily
at soft x-ray/ultraviolet wavelengths (energies $\sim 20-100$~eV).

\begin{figure}
\includegraphics[width=0.5\textwidth]{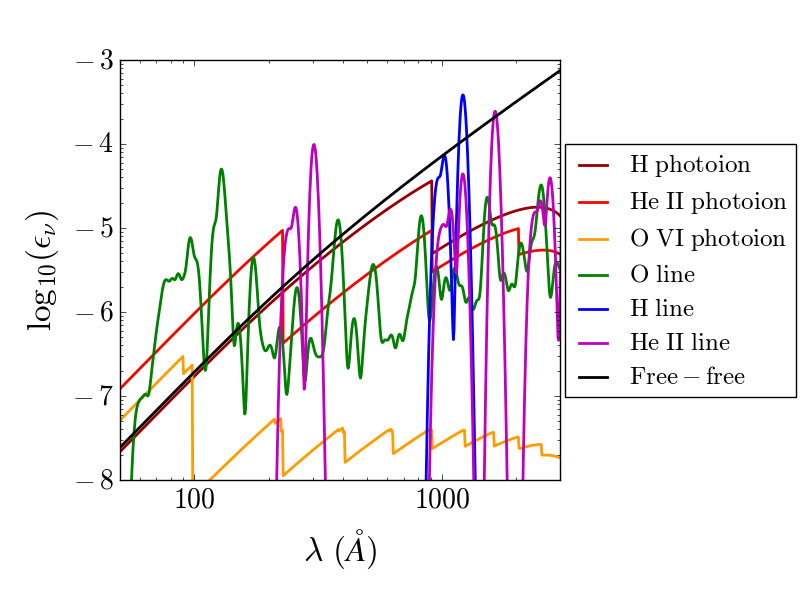}
\caption{The ratio of various absorptive opacities at soft x-ray and UV wavelengths to that
of electron scattering, computed near the inner boundary of a TDE envelope. The values are from the non-LTE radiation transport calculation of Section~\ref{sec:Results} 
with $\Menv = 0.5 \msun$, $\rin = 10^{14}$ cm, $\rout =  10^{15}$ cm, and $L = 10^{45}$ \ergss. At soft x-ray wavelengths near $100$~\AA, photoionization of He II is a dominant absorptive process. Oxygen lines also contribute, while oxygen photoionization is negligible. The photoionization opacities begin to dip  at long wavelengths due to stimulated recombination.}
\label{fig:epsilon_Xray} 
\end{figure}

Figure~\ref{fig:epsilon_Xray} shows the processes that contribute to the absorptive opacity
for conditions found in a typical numerical calculation of a TDE envelope (to be described in
Section~\ref{sec:Results}).   For soft x-rays 
near 100~\AA, a primary absorptive opacity is photoionization of HeII (threshold energy  of 54.4~eV).  Photoionization and line absorption from other metals will also contribute to the reprocessing.

 In this section, we quantify the conditions in the envelope (density, temperature, and ionization state) and the effective optical depth to
HeII photoionization.  We then estimate the  fraction of light reprocessed into the optical and its dependence on
envelope parameters.   The
results, while approximate, will be useful in interpreting the detailed numerical spectral calculations that follow.

\subsection{Envelope Density Structure and Optical Depth}
\label{sec:RadiationTemperatureProfile}
  
We invoke the presence of a reprocessing envelope surrounding the source, and remain agnostic about its origin.  
 For simplicity, we assume the envelope is spherically symmetric and  quasi-static; in reality, 
global asymmetries and  velocity gradients will complicate the picture, and are issues to be addressed in future work. 

We model the envelope with a power-law  density profile, $\rho(r) \propto r^{-p}$, extending
 from an inner radius, \rin, to an outer radius, \rout.
An exponent  $p = 3$ applies to a radiation pressure-supported envelope \citep{Loeb1997}, while $p=2$
for a steady state wind outflow (or inflow).  The hydrodynamic solutions of \citet{Coughlin2014-1} find intermediate
values of $p = 1.5$ to $3$.  In this work, we  adopt $p = 2$, in which case the density profile is
\begin{equation}
\label{eq:DensityProfile}
\rho(r) = \frac{\Menv}{4 \pi \rin^2 (\rout - \rin)} \left( \frac{r}{\rin} \right)^{-2}.
\end{equation}
The envelope mass, \Menv\ is at most the total mass of the originally bound stellar debris (generally $\lesssim 0.5~\msun$) and for a quasi-static envelope will likely change over time as material falls back, accretes or is launched in an outflow.   If the density profile of Equation~\eqref{eq:DensityProfile} is interpreted
 a steady wind,  the corresponding mass loss rate is $\dot{M} = \Menv (v_{\rm w}/\rout)$ where  $v_{\rm w}$ is the wind speed.

For quasi-static envelopes, we expect \rout\ to be set by the radiation pressure support, with typical values $\rout \sim 10^{15}~{\rm cm}$ \citep{Loeb1997, Coughlin2014-1,Guillochon2014}.   For a radiatively launched outflow,  \rout\ will be set by the extent of the expanding gas at any given time.  Gas expanding at  $v_{\rm w} = 10,000$~\kms\  for 20 days will have reached a similar radius,  $\rout \approx 10^{15}$~cm.  
Scaling to characteristic values, the density at the base of the envelope, \rin, is 
\begin{equation}
\rhoin  \approx\frac{ 8 \times 10^{-12}}{(1 - \rin/\rout)}~\Menvs \rins^{-2} \routs^{-1}~\gcc
\end{equation}
where $\Menvs = \Menv/0.5~\msun$, $\rins = \rin/10^{14}$~cm, and $\routs = \rout/10^{15}$~cm.
For solar composition, the mean particle mass is $\mu \simeq 1.3 m_{\rm p}$ and the 
number density at the base is $\nin = \rhoin/\mu \approx 4 \times 10^{12}~{\rm cm^{-3}}$.

For fully ionized gas of solar composition, the dominant continuum opacity 
is electron scattering with an opacity $\kappa = 0.32~\csg$.
The optical depth through the envelope is then
\begin{eqnarray}
  \label{eq:TotalTau}
\taues = \frac{ \kappa \Menv}{4 \pi \rin \rout} \approx 270
~\Menvs \rins^{-1} \routs^{-1}.
\end{eqnarray}
The  envelope is highly scattering dominated, with the ratio of
absorptive opacity to total opacity $\epsilon \lesssim 10^{-4}$ at x-ray/UV wavelengths
(see Figure~\ref{fig:epsilon_Xray}).
The reprocessing of radiation to optical wavelengths is accomplished via absorptive processes (e.g., HeII photoionization), 
 not electron scattering.
However, the large scattering opacity traps photons and increases the probability that they will be absorbed.  It
thus has a critical effect on the formation of the optical continuum and emission lines.    

The radiation diffusion time through the envelope is $t_{\rm diff} 
\sim  \bar{r}^2 \kappa \bar{\rho}/c$ where $\bar{\rho}$ is an appropriately weighted
envelope density and $\bar{r}$ a characteristic envelope length scale.  Taking  $\bar{\rho} = \rho(\bar{r})$ gives
$t_{\rm diff} \sim 10~\Menvs \rins^{-1}$~days, which is  similar to the diffusion
time we find in our direct numerical transport calculations of Section~\ref{sec:Results}.
For the TDE envelopes we consider here at times $t \approx$ weeks since disruption (near the peak of the light curve), $t_{\rm diff} \lesssim t$ and we can assume steady state transport above the inner boundary.
However, at earlier times on the light curve rise, or for very dense envelopes, the  diffusion timescale 
may be important in setting the
emergent luminosity.

\subsection{Envelope Temperature Structure}
\label{RadiationEnergyDensityProfile}

We assume that above our inner boundary, energy is primarily transported by radiation diffusion and
there is negligible energy input from other processes such as viscosity or shocks.  The temperature structure of the envelope is then set  by radiative heating and cooling.
Given the high electron scattering optical depth, we apply the spherical steady-state
diffusion approximation to  determine the frequency integrated radiation energy density, $E_{\rm rad}(r)$,
\begin{equation}
\label{eq:EnergyGradient}
\frac{d \Erad(r)}{dr} = - \frac{3  \kappa \rho(r)}{4 \pi c r^2} L
\end{equation}
where $c$ is the speed of light and
$L$ is the bolometric luminosity of the central source. The solution of Equation~\eqref{eq:EnergyGradient}, using the density profile from Equation~\eqref{eq:DensityProfile} and a constant electron-scattering opacity, is
\begin{equation}
\label{eq:Erad}
\Erad(r) = \frac{\taues L} {4 \pi c \rin^2}
\left[ \frac{\rin^3}{r^3} - \frac{\rin^3}{\rout^3}  + \frac{4}{\taues} \frac{\rin^2}{\rout^2} \right]
\end{equation}
where we have taken as an outer boundary condition $\Erad(\rout) = a_{\rm rad} \Teff^4$,  where $a_{\rm rad}$ is the radiation constant, $\Teff = [L/4 \pi \sigma_{\rm sb} \rout^2]^{1/4}$ and $\sigma_{\rm sb} \equiv c \, a_{\rm rad}/4$ is the Stefan-Boltzmann constant. Defining a radiation temperature by $\Trad = (\Erad/a_{\rm rad})^{1/4}$,
we have in the limit $\rout \gg \rin$ 
\begin{equation}
\Trad(r) = \Trad(\rin) \left[ \frac{\rin}{r} \right]^{3/4}~~({r \ll \rout}),
\label{eq:temp_struct}
\end{equation}
where the interior temperature is
\begin{equation}
\Trad(\rin) = \left[ \frac{\taues}{4} \frac{L}{4 \pi \sigma_{\rm sb} \rin^2} \right]^{1/4}.
\label{eq:Tin}
\end{equation}
The temperature at the inner boundary is greater than that of a blackbody sphere emitting into vacuum
by a factor of $(\tau_{\rm es}/4)^{1/4} \approx 3$.  This is due to the back-heating of the source
by photons trapped in the optically thick envelope. 
We estimate the characteristic radiation temperature to be
\begin{equation}
\Trad(\rin)  \approx 3.1 \times 10^5~L_{45}^{1/4}\Menvs^{1/4} \routs^{-1/4} \rins^{-3/4}~ {\rm K} \, \,.
\label{eq:Tscalings}
\end{equation}
Note that the mass and size of
the envelope directly affects the temperature. For simplicity, we will usually set $T_s$, the blackbody temperature of our inner source, equal to $T_{\rm rad}(r_{\rm in})$.  This may not hold in general; for example,
the source luminosity may come from an accretion disk with radius smaller than 
the inner envelope edge, \rin.

As defined, $T_{\rm rad}$ is simply a convenient rescaling of the local radiation energy density.  However, we now show that $T_{\rm rad}$ may provide a good estimate of the gas temperature \Tgas.  When the thermal state of the gas is
determined primarily by radiative heating and cooling, the  time-evolution of \Tgas\ is 
\begin{eqnarray}
  \label{eq:ElectronEnergyEquation}
  \frac{3}{2} n k_B \frac{d \Tgas}{dt} = - 4 \pi \alpha_S B(\Tgas)   +  4 \pi \alpha_E J, 
\end{eqnarray}
where $J = c \Erad /4 \pi$ is the integrated mean specific intensity, $B(\Tgas) = \sigma_{\rm sb} \Tgas^4/\pi$ is the frequency-integrated Planck function, \ $k_B$ is the Boltzmann constant, and $\alpha_E$ and $\alpha_S$ are mean absorptive extinction coefficients defined by
\begin{eqnarray}
  \label{eq:MeanOpacities}
   \alpha_E \equiv \frac{\int  \alpha_\nu^{\rm abs}(\nu) E_{\nu}(\nu) d\nu}{ E_{\rm rad} } \nonumber, ~~~~
 \alpha_S \equiv \frac{\int  \alpha_\nu^{\rm abs}(\nu) B_{\nu}(\nu) d\nu}{ B},
\end{eqnarray}
where $\alpha_\nu^{\rm abs}(\nu)$ is the absorptive extinction coefficient at each frequency, and $E_{\nu}(\nu)$ is the radiation energy density in the frequency interval between $\nu$ and $\nu + d\nu$.

Given sufficient time, the gas will reach a thermal equilibrium where radiative heating balances cooling.  From Equation~\eqref{eq:ElectronEnergyEquation},
 the  timescale, $t_{\rm heat}$, to heat gas from some lower temperature up to the equilibrium value is
 \begin{equation}
  \label{eq:HeatingTime}
  t_{\rm heat} =   \frac{3}{2} \frac{1}{ c \, \alpha_E}
  \left[ \frac{ n k_B T_{\rm rad}}{a_{\rm rad} T_{\rm rad}^4 } \right].
\end{equation}
The term in brackets is the ratio of gas to radiation energy density, and is $\ll 1$ for 
the conditions in TDE envelopes. 
The true absorption coefficient $\alpha_E$ is $\gtrsim 10^{-5}$ that of electron scattering (Figure~\ref{fig:epsilon_Xray}) from which we find $t_{\rm heat}$ ranges from $\lesssim  1$~s at the base of the envelope up to $\sim$ hours at the outermost radii, which is comfortably smaller than the characteristic timescale of weeks for the evolution of  observed TDE lightcurves.  
The gas temperature will then be able to reach a steady state, $d T_{\rm gas}/dt = 0$, and Equation~\eqref{eq:ElectronEnergyEquation} gives
\begin{equation}
  \label{eq:ElectronTempRadTempRelation}
  T_{\rm gas} = (\alpha_E/\alpha_S)^{1/4}~ T_{\rm rad}.
\end{equation}
Therefore, \Tgas\ is close to \Trad, with a correction factor related to the frequency dependence of $\alpha^{\rm abs}_\nu$ and the extent to which the radiation field spectrum differs from a blackbody.  In our numerical calculations, we find that Equation~\eqref{eq:temp_struct} provides a  good estimate of \Tgas\ in the inner portions of the envelope, up to the continuum thermalization depth. Beyond that, \Tgas\ plateaus at a higher value than \Trad.

\subsection{Envelope Ionization State}

Photoionization of HeII  provides one of the most important absorptive opacities at soft x-ray wavelengths (see Figure~\ref{fig:epsilon_Xray}).  If helium is primarily in the HeII state, the associated photoionization optical depth is
 $\gg 1$, and essentially all of the source x-rays will be absorbed and reprocessed to longer wavelengths.  
 However, for high luminosity TDEs, the intense radiation field may completely ionize helium to HeIII, allowing only a small fraction 
 of the source luminosity to be absorbed.  
 Such an  ionization effect has been explored in \cite{Metzger2014} and \cite{Metzger2015}.
 Determining the ionization state is therefore crucial for estimating 
 the reprocessing efficiency of a TDE envelope.
 
To roughly estimate the critical luminosity, \Lion, required to highly ionize the envelope, we  use simple Stromgren sphere arguments. 
The rate at which ionized photons are produced by the central
source is $\dot{Q} = \Ls/\Ep$ where \Ep\ is the average  energy of source photons.
The total recombination rate within a sphere of radius $r$ is
\begin{equation}
\dot{R}(r) = \int_{\rin}^{r} 4 \pi r'^2 \alpha_{B,0} n_e(r') n_{\rm HeIII}(r') dr'
\end{equation}
where $\alpha_{B,0} \approx 2 \times 10^{-13}~{\rm cm^3~s^{-1}}$ is the case B helium recombination coefficient at a temperature $10^5$~K, 
$n_e$ is the free electron density, and $n_{\rm HeIII}$ is the number density of HeIII.
The condition $\dot{Q} = \dot{R}(r)$ allows us to solve for the Stromgren radius 
within which the helium is fully ionized 
\begin{equation}
r_{\rm strom} = \rin \left[ 1 - \frac{L}{4 \pi A_{\rm He} \nin^2 \alpha_{B,0} \rin^3 \Ep} \right]^{-1}
\end{equation} 
where $A_{\rm He} \approx 0.1$ is the number fraction of helium, and we have assumed $n_e$ equals the ion number density, $n$, which is accurate to within 10\% for the conditions that interest us.  
We see that the Stromgren radius of HeII  diverges for luminosities above a critical luminosity
$\Lion = 4 \pi \alpha_{B,0} A_{\rm He} \nin^2 \rin^3$ or 
\begin{equation}
\Lion \approx  3 \times 10^{44}~
\Menvs^2 \rins^{-1} \routs^{-2} E_{50}~\ergss
\label{eq:Lcrit}
\end{equation}
where $E_{50} = \Ep/(50~{\rm eV})$.
A transition in the ionization state around this luminosity is confirmed by our numerical calculations in Section~\ref{sec:Results}, and has a dramatic effect on the fraction of escaping x-ray photons.
 
 Similar  arguments could
 be applied to other elements that may contribute to absorption.
However,  the Stromgren estimates are ultimately limited by the fact that, in the true radiation transport, ionizing
photons may be produced not only by the source, but also within the TDE envelope.
In particular,  radiation absorbed by helium will be largely re-emitted as photons capable
of ionizing hydrogen.  Indeed,  our numerical calculations (Section~\ref{sec:Results}) find that 
hydrogen  remains fully ionized for luminosities  less than the
 \Lion\ implied by Equation~\ref{eq:Lcrit}.

In the limit that helium is highly ionized ($L > \Lion$), we can estimate the fraction of helium in the HeII state.
Assuming that photoionization equilibrium holds at all radii, the number densities, $n_{\rm HeII}$ and $n_{\rm HeIII}$, of HeII and HeIII respectively are related by
\begin{equation}
\label{eq:IonizationBalance}
n_{\rm HeII} {\cal I} = n_{\rm HeIII} n_e \alpha_{\rm B}
\end{equation}
where ${\cal I}$ is the photoionization rate 
and the radiative recombination coefficient  depends on temperature approximately as $\alpha_B \propto T^{-1/2}$.  Using the
temperature structure from Equation~\ref{eq:temp_struct} gives
\begin{equation}
\alpha_B(r) \approx \alpha_{B,0} \Tss^{-1/2} \left[ \frac{r}{\rin} \right]^{3/8}
\end{equation}
where $\Tss = \Ts/10^5$~K.  
For typical envelope densities, collisional ionization rates are small, while the recombination timescale, $t_{\rm rec} \sim 1/n_e \alpha_B \sim 1$~s, is short, validating  the assumption of photoionization equilibrium. 
The photoionization rate is 
 \begin{equation}
  \label{eq:PhotoionrateDefinition}
{\cal I}(r) \equiv 4 \pi \int_{\nu_{\rm ion}}^{\infty} \frac{\sigma^{\rm ion}_\nu J_{\nu}(r)}{h \nu} d\nu   
\end{equation}
where \nuion\ is the threshold frequency for HeII ionization,  $J_{\nu}(r)$ is the mean specific intensity of the radiation field, and the photoionization cross-section $\sigma^{\rm ion}_\nu$ is given  
to good approximation (about 15\% error because of neglect of Gaunt factors)  for hydrogenic ions by
\begin{equation}
  \label{eq:PhotoionCS}
\sigma^{\rm ion }_{\nu}(\nu) = \sigma_{0} \left(\frac{\nu}{\nu_{\rm ion}}\right)^{-3}  
\end{equation}
with $\sigma_0 \approx 1.5 \times 10^{-18}~{\rm cm^2}$ for HeII.
In the limit that only a small fraction of the source radiation is absorbed,  the frequency dependence of $J_\nu$ will be a Planck distribution at the source temperature, $B_\nu(\Ts)$.
The radiation energy density, however, will be diluted according to
the diffusion solution Equation~\eqref{eq:Erad}, giving
\begin{equation}
J_\nu(r) = B_\nu(\Ts) \left[ \frac{\rin}{r} \right]^{3}
\label{eq:diluted_J}
\end{equation}
which assumes $\rout \gg \rin$.  The ionization rate is then
\begin{equation}
{\cal I}(r) = \frac {8 \pi \nu_{\rm ion}^3 \sigma_0}{c^2}\left[ \frac{\rin}{r} \right]^{3} \Omega(\Ts)
\end{equation}
where the dimensionless factor $\Omega$ is 
\begin{equation}
\Omega(\Ts) = \int_{\frac{h \nuion}{k_B \Ts}}^\infty \frac{dx}{x(e^{x} - 1)} \approx 0.8
\left[ \frac{k_B \Ts}{h \nuion} \right]
e^{-h \nu_{\rm ion}/k_B \Ts}.
\label{eq:Omega}
\end{equation} 
The second expression  approximates the integral to within a few percent over the range of temperatures of interest.  
The ionization equilibrium Equation~\eqref{eq:IonizationBalance} then determines the fractional ratio of HeII 
\begin{equation}
f_{\rm HeII} = \frac{n_{\rm HeII}}{n_{\rm HeIII}} = \left[ \frac{ n_{\rm in} \alpha_{B,0} c^2}{8 \pi \nu_{\rm ion}^3 \sigma_0 \Tss^{1/2} \Omega(\Ts)} \right] 
\left[\frac{r}{\rin} \right]^{11/8}
\label{eq:ion_state}
\end{equation}
where we have again taken $n_e = n$. The fraction of HeII grows with radius, due both to the decrease of the ionizing radiation field and the increase of the recombination coefficient at
larger radii. 
The HeII fraction at \rin\ is
\begin{equation}
f_{\rm HeII}(\rin) = 6 \times 10^{-11} e^{\frac{6.31}{ \Tss}}
  \Tss^{-3/2} \Menvs \rins^{-2} \routs^{-1}
\label{Eq:ionization}
\end{equation}
which shows that, for these  fiducial parameters, most of the helium is in the fully ionized HeIII state.  This predicted  HeII fraction is very similar to what we find in our
 numerical transport calculations (Section~\ref{sec:Results}) when $L \gg \Lion$.  
The temperature dependence in Equation~\ref{Eq:ionization} resembles the LTE Saha equation expression,
a consequence of the assumed Planckian frequency distribution of the radiation field.  Near the inner boundary, where the radiation field approaches a true blackbody, the ionization
state will approach its LTE value.

\subsection{Reprocessed Luminosity}

Having solved for the ionization state, we can calculate the optical depth to HeII photoionization 
\begin{equation}
\tau_{\rm HeII}(\nu) = \int_{\rin}^{\rout} n_{\rm HeII}(r) \sigma_0 (\nu/\nuion)^{-3} dr.
\end{equation}
Figure~\ref{fig:epsilon_Xray}  makes it clear that other  opacities (e.g., HeII and oxygen lines) also contribute to the opacity soft x-ray wavelengths;  however, our analysis
of the HeII photoionization provides some proxy for the more general
radiative processes.

When most of the helium is in the HeIII state (as indicated by Equation~\ref{Eq:ionization}) we can write the number density of HeII as 
$n_{\rm HeII} = A_{\rm He} f_{\rm HeII} n$.  
For  the temperature range of interest ($2 \times 10^5~{\rm K} < \Ts < 10^6~{\rm K}$) we 
can more coarsely approximate Equation~\eqref{eq:Omega} as $\Omega \approx 6 \times 10^{-3} \Tss^{2}$. 
Using our ionization solution Equation~\eqref{eq:ion_state}
gives
\begin{align}
&\tau_{\rm HeII}(\nu) = \nonumber \\
&\frac{A_{\rm He} \nin^2 \rin \alpha_{B,0} c^2}{3 \pi \nu_0^3 \Tss^{1/2} \Omega(T_s)} \left[ \frac{\nu}{\nuion} \right]^{-3} \left[ \left(\frac{\rout}{\rin}\right)^{3/8} - 1 \right] \\
&\approx 0.02 ~L_{45}^{-5/8}  \Menvs^{11/8} \rins^{-3/2} \routs^{-1} (\nu/\nuion)^{-3}.
 \end{align}
The expected fraction of absorptive to electron scattering opacity at threshold is $A_{\rm He} f_{\rm HeII}(\rin) \sigma_0/\sigma_T \approx 2 \times 10^{-5}$ (where $\sigma_T$ is the Thomson cross-section for electron scattering), in rough agreement with our numerical results (Figure~\ref{fig:epsilon_Xray}).

In the radial direction, the envelope is optically thin to HeII photoionization.  However, the
electron scattering opacity increases the path length of photons as they random walk through the envelope,
enhancing the probability of absorption.   The typical number of scatters
is $\taues^2$, and so the effective optical depth
for a photon to be absorbed is $\tau_{\rm r} = \sqrt{ \tau_{\rm HeII} \taues} $, or
\begin{equation}
\tau_{\rm r}(\nu) \approx 2 L_{45}^{-5/16} \Menvs^{19/16} \rins^{-5/4} \routs^{-1}   (\nu/\nuion)^{-3/2}
\label{eq:tau_r}
\end{equation}
For frequencies near the HeII threshold, the envelope
can thus absorb and re-emit a fraction of the
source luminosity, \Ls. 
As a simple estimate of the reprocessed luminosity, 
we assume that a fraction $e^{-\tau_{\rm r}}$ of the source luminosity is 
absorbed and re-emitted as a blackbody
at an envelope temperature \Tr.  The specific
luminosity of reprocessed light is then
\begin{equation}
L_{\rm \nu}(\nu) = e^{-\tau_{\rm r}}  \Ls \frac{B_\nu(\Tr)}{\int_0^\infty B_\nu(\Tr) d \nu}.
\end{equation}
If observations are taken at a frequency $\vopt = 4000$~\AA\ that is on the Rayleigh-Jeans tail of
the blackbody ($h \vopt \ll k_B \Tr$), and when  $\tau_{\rm r} \ll 1$
the observed luminosity is  \begin{eqnarray}
\vopt L_{\rm \nu}(\vopt) =  \vopt \tau_{\rm r}  \Ls 
\left[ \frac{2 \pi \vopt^2 k_B }{c^2 \sigma_{\rm sb} \Tr^3}\right] \\
\approx
1.1 \times 10^{43}~
 L_{45}^{-1/16}  \Menvs^{7/16}\rins \routs^{-1/4}~\ergss.
 \label{eq:L_opt}
\end{eqnarray}
The normalization in the second expression depends on the assumed temperature
\Tr\ of the reprocessed radiation; the value in Equation~\ref{eq:L_opt} assumes
is emitted a radius of $5 \rin$ where   $\Tr \approx \Ts/3$.

Though  approximate, our analytic treatment provides insight into how
reprocessing  takes place in highly ionized TDE envelopes. 
The reprocessed optical luminosity increases with \Menv, though sub-linearly.
Having more mass in the envelope increases the effective absorptive optical depth,  but there is also a counteracting effect; 
a higher \Menv\  produces higher envelope and source temperatures,  which increases the ionization
state and shifts the reprocessed emission to shorter wavelengths. 

In the highly ionized regime, the reprocessed optical luminosity depends very weakly on the source luminosity.   This is because a higher
\Ls\ leads to higher ionization state, and hence a lower  HeII effective optical depth;
although the input radiation is brighter, a smaller fraction of it is absorbed and reprocessed.
This behavior, however, only holds in the  limit that a small fraction of the  source x-rays are absorbed. When $L \lesssim \Lion$, 
HeIII will recombine and a HeII ionization front will develop in the envelope.   In this case, nearly all of the source  photons with energies $\gtrsim 54.4$~eV are absorbed.  The reprocessing  becomes highly efficient and the optical luminosity will more closely track the input luminosity, in contrast to the weak dependence found in Equation~\eqref{eq:L_opt}.

The scalings in Equation~(\ref{eq:L_opt}) also do not reflect the likely correlation between  the parameters. For example, both the envelope mass and bolometric luminosity may be decreasing at the light curve peak. In this situation, the bolometric luminosity and the reprocessed luminosity will track each other more closely than what Equation~(\ref{eq:L_opt}) predicts when  $L$ is varied on its own.

\subsection{Optical Line and Continuum Formation}

The soft x-rays absorbed by photoionization in the envelope can be re-emitted
at longer wavelengths.  In an optically thin medium (e.g., an
HII region) absorbed photons are  primarily re-emitted in  lines.  In TDE envelopes, in contrast, the medium can be effectively optically thick in the continuum, and radiation will be reprocessed  into both continuum and lines.

\begin{figure}
\includegraphics[width=0.5\textwidth]{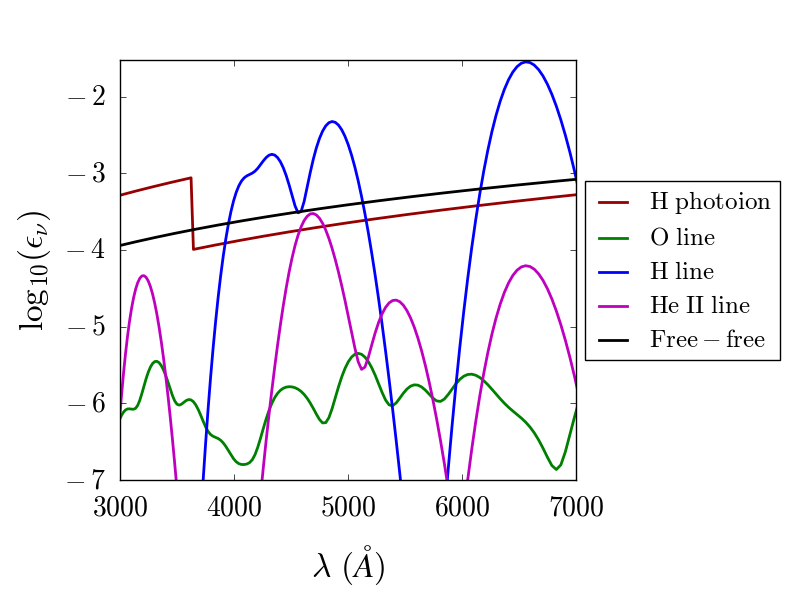}
\caption{The ratio of various absorptive opacities at optical wavelengths to that
of electron scattering, computed near the outer edge of a TDE envelope.  The values are from the non-LTE radiation transport calculations of Section~\ref{sec:Results} with 
$\Menv =  0.5 \msun$, $\rin = 10^{14}$ cm,  $\rout = 10^{15}$ cm, and $L = 10^{45}$ \ergss. Free-free is the dominant continuum opacity at wavelengths longer than the Balmer break.  Hydrogen Balmer series lines provide the highest opacities of all, with \Ha\ being the most opaque. Oxygen lines provide negligible opacity. 
\label{fig:epsilon}}
\end{figure}

Figure~\ref{fig:epsilon} shows an example of the relevant opacities for TDE envelopes at optical wavelengths from a numerical calculation to be discussed in Section~\ref{sec:Results}.  The primary continuum opacity
is free-free (bremsstrahlung), with some contribution from
 bound-free absorption from excited states of hydrogen and helium.
The free-free absorption coefficient in the Rayleigh-Jeans limit ($h \nu \ll k_B T$) is
\begin{equation}
\alpha_{\rm ff} \approx 2 \times 10^{-2} n~n_e T^{-3/2} g_{\rm ff} \nu^{-2}
\end{equation}
where the free-free Gaunt factor,  $g_{\rm ff}$, is of order unity.  This can be compared
to the electron scattering absorption coefficient $\alpha_{\rm es} = n_e \sigma_t$ to
give the absorption ratio $\epsilon_{\rm ff} = \alpha_{\rm ff}/(\alpha_{\rm ff} + \alpha_{\rm es}) \approx \alpha_{\rm ff}/\alpha_{\rm es}$ of
\begin{equation}
\epsilon_{\rm ff} \approx 10^{-3} L_{45}^{-3/8} \Menvs^{5/8} \rins^{-7/8} \routs^{-5/8} \left[ \frac{\nu}{\vopt} \right]^{-2}
\end{equation}
which has only a weak dependence on radius.   
The radial optical depth to free-free is $\tau_{\rm ff} = \epsilon_{\rm ff} \taues < 1$.  However, the multiple electron scattering enhances the effective optical
depth of free-free absorption. 
The continuum emission originates roughly from the thermalization depth to free-free absorption at an electron scattering optical depth of  $\tau_{\rm therm} = 1/\sqrt{\epsilon} \approx 30$.

Some optical lines may also be effectively optically thick. The absorption coefficient for a bound-bound transition between and lower and upper level with number densities $n_l$ and $n_u$ is
\begin{equation}
  \label{eq:BoundBound}
  \alpha_{\rm bb} = \left( \frac{ \pi e^2}{m_e c} \right) n_l f_{\rm osc}  \phi_{\nu} (\nu) \left[ 1 - \frac{g_l n_u }{g_u n_l}\right]
\end{equation}
where $f_{\rm osc}$ is the oscillator strength and $\phi_\nu$ the line profile function.  The term in brackets is the correction for stimulated emission 
where $g_l$ and $g_u$ are the statistical weights.   To estimate the extinction
at the line center frequency, $\nu_0$, we approximate $\phi(\nu_0) \approx 1/\Delta \nu$,
where for a line Doppler broadened by a velocity $v$ the line width $\Delta \nu = \nu_0 (v/c)$. 
 When the radiation field takes the form of a diluted
blackbody (Equation~\ref{eq:diluted_J}), the relative level populations will approximate their LTE values, e.g., $n_u/n_l \approx (g_u/g_l) e^{-\Delta E_l/k \Ts}$.  In the limit $h \nu_0 \ll k \Ts$,
Equation~\ref{eq:BoundBound} then becomes
 \begin{equation}
 \alpha_{\rm bb,0} \approx  \left( \frac{ \pi e^2}{m_e c} \right) n_l f_{\rm osc} \left[ \frac{c}{v}  \right] \frac{h}{k_B\Ts}.
 \end{equation}
 The ratio $\epsilon_{\rm bb} =  \alpha_{\rm bb}/\alpha_{\rm es}$ is then
 \begin{equation}
 \epsilon_{\rm bb} \approx 0.06 \frac{n_l}{10^{-10} \nin} f_{\rm osc} v_9^{-1} \Tss^{-1}
 \end{equation}
 where $v_9 = v/10^9~{\rm cm~s^{-1}}$ and the density $n_l$ is scaled by
 the expected ionization fraction of Equation~\ref{Eq:ionization}.
The \Ha\ line emission thus originates from a region of electron
scattering depth of $\tau_{\rm therm} = 1/\sqrt{\epsilon} \approx 6$.
Due to the lower abundance of helium, $A_{\rm He} \approx 0.1$, the optical depth of the HeII~$\lambda4686$ line (all other things being roughly equal) should be smaller than \Ha\ by
a factor $\sim 10$,
suggesting a larger thermalization depth $\tau_{\rm therm} \approx 20$.
The values roughly agree with the numerical results of Figure~\ref{fig:epsilon}.

These arguments motivate the physical picture illustrated in the schematic of Figure~\ref{fig:schematic}.  In this quasi-static picture,  spectrum formation resembles that of a stellar atmosphere,
with the emission at different wavelengths originating from the source function at different thermalization depths. 
Soft x-rays  are generated in the interior (at our inner boundary).  The optical continuum forms further out, while the effective photospheres of strong lines of hydrogen and helium  lie even nearer the surface.
All of  the observed emission is generated below the electron
scattering photosphere.  

\section{Numerical Results}
\label{sec:Results}

In this section, we present  synthetic non-LTE spectra for TDE envelopes, discussing the physics of optical line and continuum formation and the dependence on envelope
parameters.  Details of the numerical method and radiative processes treated
are given in the Appendix.    In all models, a luminosity $L$ is emitted 
as a blackbody of temperature \Ts\ at the absorbing inner  boundary \rin, and transported
through the spherical envelope of mass \Menv\ with a density structure
$\rho(r) \propto r^{-2}$ extending from radii \rin\ to \rout. In order to avoid an artificially abrupt truncation of the envelope, we allow the envelope extend beyond \rout\ following an $r^{-10}$ density profile (we find that our results are not highly sensitive to the exact value of this power-law).
We include opacities from hydrogen, helium, and oxygen,
in solar abundances.  

For the bound-bound transitions, we assume a Gaussian line profile with a  Doppler velocity of $10^4$ km s$^{-1}$.  This velocity, which  is motivated by the width of line features observed in TDE candidate spectra \citep[e.g.,][]{Gezari2012,Arcavi2014}, is much higher than the ion sound speed, but comparable to the virial velocity in TDE envelopes.  Our setup thus resembles a quasi-static envelope with disordered  velocities due to, e.g., turbulence or irregular motion driven by fallback streams, as seen in numerical simulations \citep{Ramirez-Ruiz2009, Guillochon2014}.
In envelopes
with ordered bulk velocity 
due to, e.g.,  outflow or rotation, the line formation may differ  from that discussed here.
Future calculations will consider more general velocity structures, and include a more complete metal composition.

\subsection{Spectral Energy Distributions}
\label{sec:SEDs}

\begin{figure}
\includegraphics[width=0.5\textwidth]{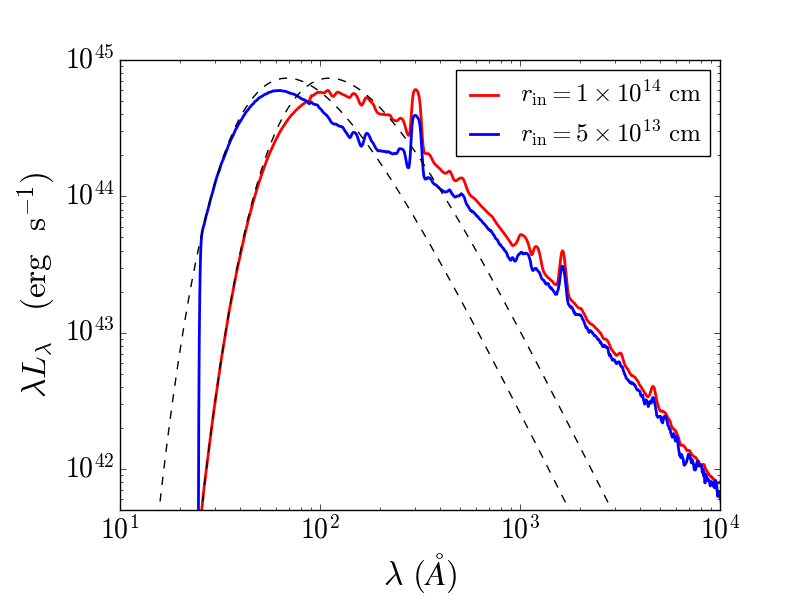}
\caption{Model spectra for two calculations with different inner boundary conditions. The two dashed curves represent the input radiation fields at the inner boundaries, with temperatures set by Equation~\eqref{eq:Tin}. In both cases, a fraction of the intrinsic spectrum is absorbed and re-radiated at UV and optical wavelengths;  the colored curves show the SED of the radiation that escapes.  In both  calculations, $\Menv = 0.25~\msun$, $ \rout = 5 \times 10^{14}$ cm (not including the $\rho \propto r^{-10}$ density tail), and $L = 10^{45}~\ergss$. 
Changing \rin\ shifts the peak of the escaping x-ray emission, but leads to only small changes in the optical emission.
\label{fig:SEDVaryInnerRadius}
}
\end{figure}

Figure~\ref{fig:SEDVaryInnerRadius} shows computed spectral energy distributions (SEDs) for two models with 
 $\Menv = 0.25~\msun$ and  $L = 10^{45}~\ergss$ ($\approx L_{\rm Edd}$ of a $10^7~\msun$ BH) but with two different values of the inner boundary
radius. The dashed curves show the unprocessed blackbody spectrum  from the inner boundary.  We find that reprocessing of the source light by the TDE envelope  enhances the optical
luminosity by several orders of magnitude.  The presence of the extended envelope  is clearly required to approach the optical luminosity of
$\sim 10^{43}~\ergss$ observed in TDE candidates.

As suggested by our analytic estimates (Equation~\ref{eq:tau_r}) for the chosen values
of $L$ and \Menv, most of the soft x-ray source emission is able to propagate through the envelope without being absorbed. Those photons that are absorbed are mostly used to photoionize HeII at wavelengths near 220~\AA.  For the $\rin = 5 \times 10^{13}$~cm model, oxygen absorption at the shortest wavelengths also plays a role.  The radiation absorbed by photoionization (though small) is re-emitted over a wide range of longer wavelengths and significantly enhances the optical flux. The simultaneous emission of both soft x-rays and optical light with an SED that does not follow a single blackbody bears a qualitative resemblance to PTF10iya, ASASSN-14li, and ASASSN-15oi \citep{Cenko2012-1,Holoien2016-1,Miller2015-2,Cenko2016,Holoien2016-2}. 

The reprocessed optical continuum in Figure~\ref{fig:SEDVaryInnerRadius} is relatively insensitive to the chosen inner radius, and follows a power law that superficially resembles the Rayleigh-Jeans tail of a blackbody.  However, the emission is not that of a single blackbody;  the thermalization depth varies with wavelength, and the continuum is the superposition
of thermal emission from different temperature blackbodies at different radii.  Due to this effect, the slope of our model continuum, $L_\lambda \propto \lambda^{-3}$, is somewhat shallower than that of a true Rayleigh Jeans tail, $L_\lambda \propto \lambda^{-4}$.

Given the non-blackbody nature of the emergent spectrum, the turn-over that one begins to see at bluer wavelengths cannot necessarily be extrapolated with a Planck function. This suggests caution when inferring a bolometric luminosity by fitting a single blackbody temperature to the optical continuum. In both models of Figure~\ref{fig:SEDVaryInnerRadius}, the bolometric luminosity is $10^{45}$ erg s$^{-1}$,  much larger than one might estimate based on fitting a single blackbody to the optical/UV data. 

\begin{figure}
\includegraphics[width=0.5\textwidth]{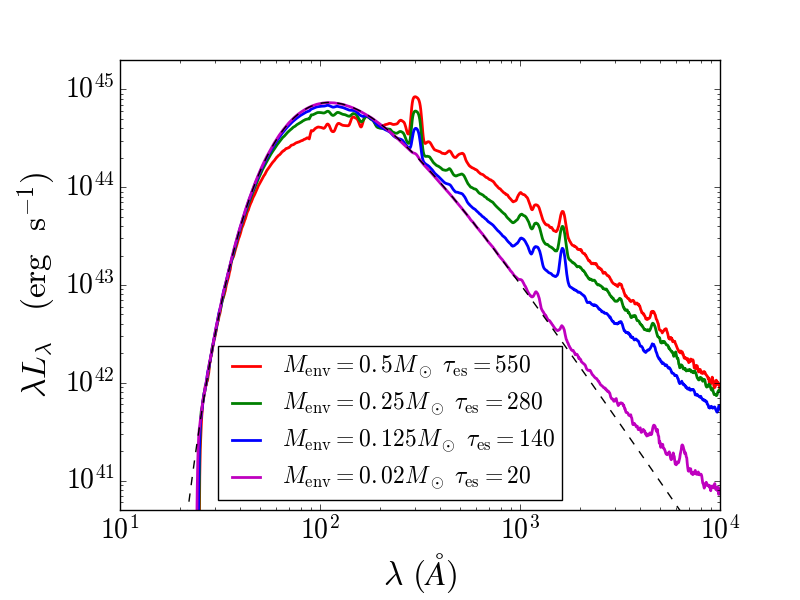}
\caption{Model spectra for calculations with varying mass in the reprocessing envelope. In all models, $L = 10^{45}~\ergss$, $\rin = 10^{14}$ cm, $\rout = 5 \times 10^{14}$ cm (not including the $\rho \propto r^{-10}$ density tail, which contributes negligible mass), and the inner boundary temperature is held fixed at $3.29 \times 10^5$ K.
Varying \Menv\ changes the amount of reprocessed optical emission in a manner similar to the prediction from Equation~\eqref{eq:L_opt}. Such behavior could represent the time-dependent depletion of the envelope depletion after peak light. This variation leaves the shape of the optical continuum mostly unchanged, at least until \Menv\ drops so low that barely any radiation is reprocessed. 
}
\label{fig:SEDVaryMass} 
\end{figure}

Figure~\ref{fig:SEDVaryMass} shows how the model SEDs depend on the envelope mass, \Menv. For this figure, $\rin = 10^{14}$~cm and for clarity we held the temperature of the inner boundary emission fixed at $3.29 \times 10^{5}$ K (unlike the rest of the models we present in this work, we did not set \Ts\ equal to $T(\rin)$ as given by Equation~\ref{eq:Tin}).  As expected from our analytic arguments (Equation~\ref{eq:L_opt}) the optical luminosity increases with \Menv\ due to the greater degree
reprocessing by  higher mass envelopes. 
 For the smallest envelope mass ($\Menv = 0.02$ $M_\odot$) essentially
none of the source luminosity is reprocessed, despite the fact that the gas is optically thick
to electron scattering ($\tau_{\rm es} \approx 20$).  The  emphasizes that in
scattering dominated TDE  envelopes, an optical depth $\gg 1$ is required to thermalize and reprocess x-rays to optical wavelengths.

\begin{figure}
\includegraphics[width=0.5\textwidth]{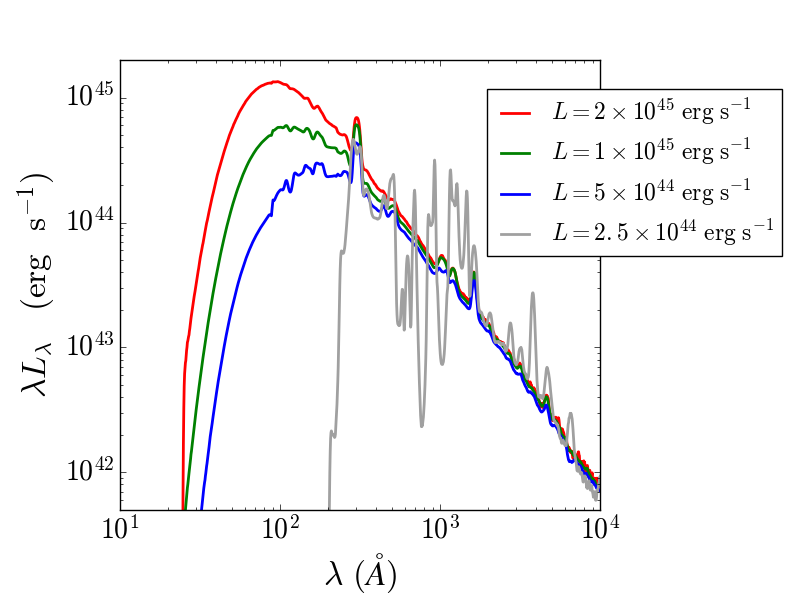}
\caption{Model spectra for calculations with varying bolometric luminosity. In all models,  $\Menv = 0.25 \msun$, $\rin = 10^{14}$ cm and $\rout = 5 \times 10^{14}$ cm (not including the $\rho \propto r^-{10}$ density tail). Varying the bolometric luminosity alone has only a minor effect on the emitted optical flux, as predicted in Equation~\eqref{eq:L_opt}. The pattern changes when the luminosity drops low enough that a helium recombination front begins to form at $L \lesssim 3\times 10^{44}~\ergss$.
 \label{fig:SEDVaryLuminosity}}
\end{figure}

Figure~\ref{fig:SEDVaryLuminosity}  shows how the model SEDs depend on the source
 luminosity, $L$.  Interestingly, the optical continuum luminosity  remains largely
 unchanged even as $L$ is varied by a factor of $\sim 10$ (again, holding the mass and size of the envelope fixed). This is consistent with
the weak $L$ dependence found in our analytic arguments (Equation~\ref{eq:L_opt}) and reflects a self-regulating  effect in the radiation transport.  Higher $L$ leads to a higher ionization state
and higher envelope temperatures, which reduces the fraction of x-ray emission that is reprocessed to longer wavelengths.

A dramatic change in the spectra of  Figure~\ref{fig:SEDVaryLuminosity} is seen when the source luminosity is reduced to $2.5 \times 10^{44}~\ergss$.
This is due to the formation
of a HeII recombination front that absorbs essentially all radiation at wavelengths
below the photoionization threshold ($\lesssim 220$~\AA).  This transition occurs roughly near the critical luminosity estimated by Stromgren arguments
in Equation~\eqref{eq:Lcrit}.  For these lower luminosities, nearly all of the radiation emerges at UV/optical wavelengths, with essentially no flux escaping 
in x-ray bands.  In this regime, the UV/optical luminosity will track the
source luminosity, in contrast to the weak $L$ dependence found in the fully ionized
regime.

\subsection{Spectral Line Features}
\label{sec:Spectra}

The spectra of Figures~\ref{fig:SEDVaryInnerRadius}--\ref{fig:SEDVaryLuminosity} show a number of  line features superimposed on the continuum.  In the soft x-ray/UV bands, the strongest lines
  are those of the He II Lyman series at wavelengths between 200 and 400~\AA. 
  Other UV lines from highly ionized species  might appear if more metals had
  been included in our calculations.
   In the optical bands, lines of hydrogen (the Balmer series) and HeII (the $4686$~\AA\  and $3203$~\AA\ lines) may be visible. We find that the line corresponding to the $n = 7$ to $n = 4$ transition in He II with wavelength $5412$~\AA\ (a Pickering series line, analogous to Brackett~$\gamma$ in H) appears for some of our models. Finally, we note that the He II Pickering line at $6560$~\AA\ has a small contribution to emission and opacity near \Ha, as pointed out in \citet{Gaskell2014}.

\begin{figure*}
\includegraphics[width=\textwidth]{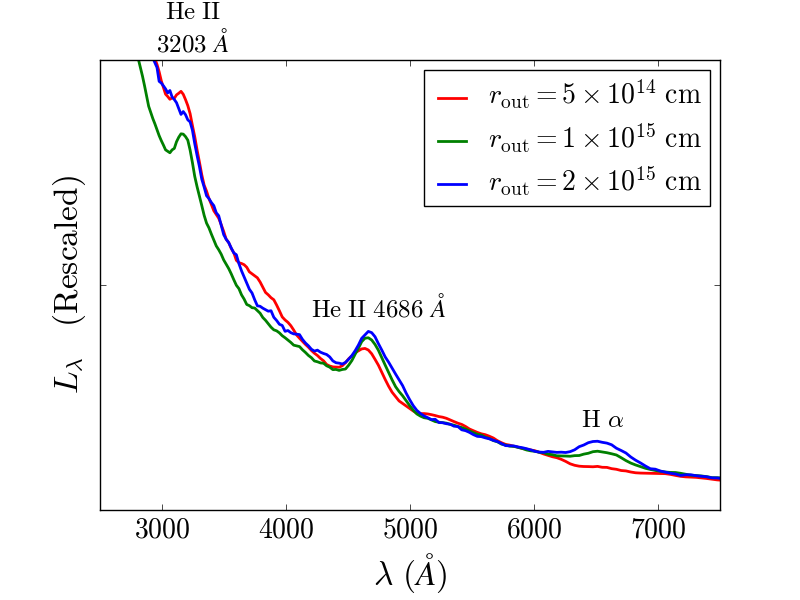}
\caption{Optical spectra for three calculations with varying outer radii for the reprocessing envelope.
 All models have $\Menv = 0.25~ \msun$, $\rin = 10^{14}$ cm, and $L = 10^{45}~ \ergss$. The flux units have been rescaled so that the continua of all three calculations overlap at $6000$ $\AA$. Two emission  lines of HeII
are visible at $3203$~\AA\ and $4686$~\AA. The \Ha\ equivalent width decreases as the envelope is made more compact. All calculations used 1600 logarithmically spaced wavelength bins between $10$ and $10^5$ $\AA$. In order to suppress the Monte Carlo noise, we have smoothed the data here to an effective resolution of 400 bins.  N.B., the indicated outer radius does not include the $\rho \propto r^{-10}$ tail.
\label{fig:SpectraVaryOuterRadius}}
\end{figure*}

\begin{figure}
\includegraphics[width=0.5\textwidth]{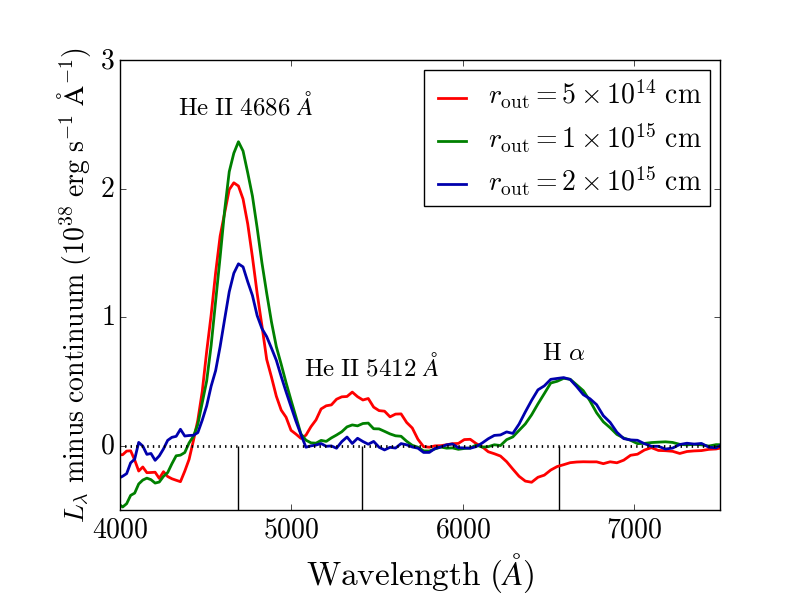}
\caption{Continuum-subtracted spectra for the same three calculations shown in Figure~\ref{fig:SpectraVaryOuterRadius}. We have subtracted off a power-law continuum to facilitate comparison of the line strengths. The helium-to-hydrogen line ratio is approximately 3:1 for the calculation with  $\rout = 2\times 10^{15}$ cm, and is 5:1 for the calculation with  $\rout = 1\times 10^{15}$ cm. For the calculation with  $\rout = 5\times 10^{14}$ cm, \Ha\ has transitioned to a shallow absorption feature. The vertical black lines at the bottom of the plot indicate the line-center wavelengths of the three labeled spectral features.}
\label{fig:SpectraContinuumSubtracted}
\end{figure}

The relative strength of the optical hydrogen and helium lines has generated particular interest in the literature, as this bears on the gas composition 
and hence the nature of the disrupted star.
Figure~\ref{fig:SpectraVaryOuterRadius} shows the optical spectra for three models
with $L = 10^{45}~\ergss$, $\Menv = 0.25~M_\odot$,
but  different values for  the outer radius of the envelope, \rout. 
In all cases, the emission in the HeII~$\lambda4686$ line exceeds that of \Ha, despite
the gas having solar composition.
As \rout\ is decreased, the \Ha\ emission decreases with respect to the continuum. 
The  line ratios can be seen more clearly in Figure~\ref{fig:SpectraContinuumSubtracted}, in which we have subtracted off a power-law to approximate the underlying continuum. The hydrogen-to-helium line ratios are roughly 3:1 and 5:1 for \rout\ of $2 \times 10^{15}$ and $1 \times 10^{15}$, respectively. For $\rout = 5 \times 10^{14}$, the \Ha\ feature has transitioned into a shallow absorption.

Our calculations thus lend further support to the interpretation that the absence of a conspicuous \Ha\ feature -- as 
observed in PS1-10jh and PTF-09ge -- 
is consistent with the disruption of a main-sequence star of solar composition.  In general, TDE envelopes
can produce a  range of \Ha\ equivalent widths and hydrogen-to-helium line ratios, depending on the gas configuration and bolometric luminosity.  Such a variation might help to explain the diversity of hydrogen-to-helium line ratios seen in observed TDE candidates \citep{Arcavi2014}.

For lower  luminosities ($L < \Lion$) recombination fronts of helium,
and eventually hydrogen  form in the envelope.  
The recombination fronts are generally accompanied by an increase in the
strength of spectral features (see Figure~\ref{fig:SEDVaryLuminosity}), including 
lines of hydrogen, helium, and oxygen.
A more  featureless spectrum is more easily obtained
when the ionization is high, which
for  lower luminosities ($L \lesssim 10^{44}~\ergss$) may require
a correspondingly smaller envelope mass or a larger outer radius.

\subsection{Understanding the Line Ratios}
\label{sec:EmissivityExplanation}

The origin of the low \Ha\ to HeII emission line ratio found in our models is due to our inclusion of optically thick radiation transport effects.  

In a TDE envelope, emitted line photons do not escape straightaway, but rather random walk through multiple
electron scatters and are subject to re-absorption  either in the line itself, or by continuum processes.  This can significantly alter the line emission as compared to the optically thin assumption.

An observer will primarily see radiation at a given wavelength emitted from the volume where the electron scattering optical depth (integrated from the outside in) is less than 
the thermalization depth $\tau_{\rm therm} = 1/\sqrt{ \epsilon }$ for that wavelength.  Since this is only an estimate, there is some flexibility in the definition of $\tau_{\rm therm}$, especially because $\epsilon$ varies with radius, along with the wavelength variation displayed in Figure~\ref{fig:epsilon}.  Once we have a thermalization depth, we can estimate the observed specific luminosity at any given wavelength by integrating the emissivity, $j_\lambda$, outside of the thermalization depth
\begin{eqnarray}
\label{eq:SpectralFeaturePrediction}
L_\lambda
&\approx&  \int_{r_{\rm therm}}^{{\infty}}  4 \pi j_\lambda  (4 \pi r^2) dr .
\end{eqnarray}
The emissivity in Equation~\eqref{eq:SpectralFeaturePrediction} is a sum over both line and continuum processes, but not electron scattering. 

 Note that photons absorbed in a line have some probability of either being line-scattered (re-emitted in the same bound-bound transition), or destroyed via multiple mechanisms (e.g. the atom de-exciting through a different line transition, being collisionally de-excited, being photoionized, or being radiatively excited to another bound electron level). The line contribution to the emissivity $j_\lambda$ in Equation~\eqref{eq:SpectralFeaturePrediction} accounts for all of these possibilities; it depends on the level populations of the bound electrons, which is in turn governed by balancing the transition rates between all bound levels and the continuum (see Appendix). 

Figure~\ref{fig:emissivity_no_line} applies this estimate of $L_\lambda$ to the calculation corresponding to the red curve in Figure~\ref{fig:SpectraContinuumSubtracted} (with $\rout = 5 \times 10^{14}$~cm).  We have plotted the integrated emissivity (Equation~\eqref{eq:SpectralFeaturePrediction}) at the wavelengths corresponding to \Ha\ and He II~$\lambda4686$.

\begin{figure}
\includegraphics[width=0.5\textwidth]{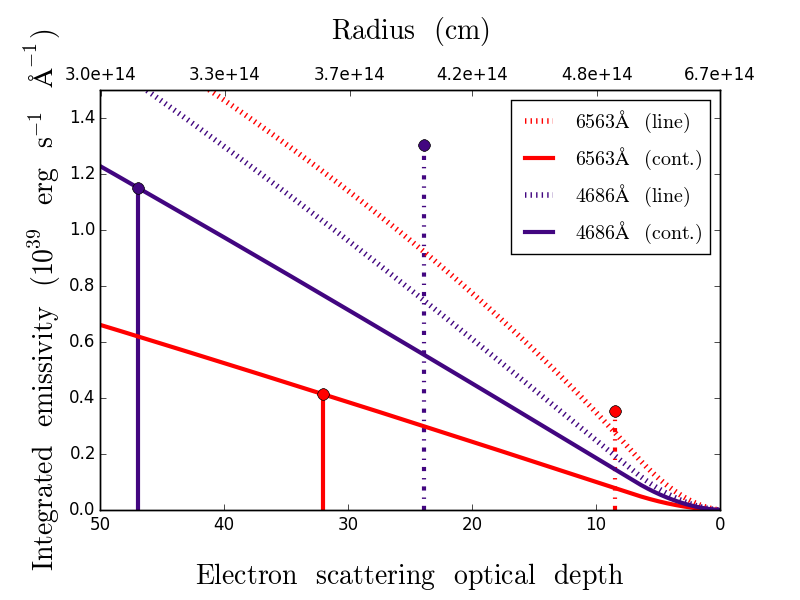}
\caption{Volume-integrated emissivity as a function of optical depth. The integrated emissivity is that expressed in equation \eqref{eq:SpectralFeaturePrediction}. The vertical lines represent the approximate thermalization depths at the indicated wavelength, either accounting for both line and continuum opacities (the dash-dotted vertical lines), or only the continuum opacities (the solid vertical lines). The plotted points represent the emission as estimated by the integrated emissivity down to the the thermalization depth, either accounting for emission from both line and continuum processes (the points on top of the dash-dotted vertical lines), or from the continuum alone (the points on top of the solid vertical lines). The line-plus-continuum emission at the \Ha\ wavelength lies very close to the emission from the continuum on its own. In contrast, the line-plus-continuum emission at $4686$ $\AA$ lies above the emission from the continuum on its own at that wavelength.}
\label{fig:emissivity_no_line}
\end{figure}

For each wavelength, we have separated the contributions from continuuum processes (the solid curves) and from lines (the dashed curves). We have indicated the approximate location of the thermalization depth, including contributions from both line and continuum opacity, with vertical dash-dotted lines. Meanwhile, thermalization depths that account for only continuum opacities are indicated with solid vertical lines. To compute the thermalization depths, we have evaluated $\epsilon$ at \rout, in this case $5 \times 10^{14}$~cm. By summing the contributions of the integrated emissivity for both the lines and the continuum processes at the thermalization depth, we obtain an estimate for the luminosity observed at each wavelength. These estimates are represented by the points plotted on top of the dash-dotted vertical lines. For reference, we have also estimated the emission from the continuum alone at these wavelengths by evaluating the integrated continuum-only emissivity at the continuum-only thermalization depth. These reference luminosities are plotted as points on top of the solid vertical lines.

As suggested by our analytic arguments (Section~\ref{sec:analytics}), the thermalization depth at the \Ha\ wavelength (the red dash-dotted vertical line) lies within several electron scattering mean free paths from the Thomson photosphere, while HeII~$\lambda4686$ has a thermalization depth several times deeper (the purple dash-dotted line).  The thermalization depths of the continuum alone at optical wavelengths are deeper still (the solid vertical lines). Even though the \Ha\ emissivity is greater at each radius than the emissivity in the HeII line, the self-absorption of line photons reduces the effective emitting volume.  The end result is that the total emission at the \Ha\ wavelength lies very close to the emission from the continuum alone at that wavelength. In contrast, the total emission at 4686 $\AA$ lies above the emission from the continuum alone at that wavelength.

The method described above provides insight into the relative strengths of the line and continuum emission at each wavelength, once the envelope ionization state and bound electron level populations are known. More accurate values for the  emission are provided directly from the Monte Carlo radiative transfer, which is why the estimates for the emission in Figure~\ref{fig:emissivity_no_line} differ slightly from what is shown in Figure~\ref{fig:SpectraContinuumSubtracted}.

Figure~\ref{fig:emissivity_no_line} makes it clear that the emergent \Ha\ feature is sensitive to the particular model parameters, which set the exact location of $r_{\rm therm}$ and  the line emissivity above it.  For different envelope configurations,  the \Ha\ line will therefore show different levels of emission, as we found in
our synthetic spectra (Figure~\ref{fig:SpectraVaryOuterRadius}).  For TDEs from massive BHs and near peak brightness, however, we typically expect a situation similar to that of Figure~\ref{fig:emissivity_no_line}, where the \Ha\ emission
is suppressed  relative to that of HeII. 

Our analysis here can be distinguished from previous studies of emission line ratios in TDEs. \cite{Guillochon2014}  suggested that \Ha\ emission in the inner layers could be suppressed by a strong ionizing continuum, and that the line strength would thus correlate with the spatial extent of solar composition gas. This conclusion, however, was based on an interpretation of existing CLOUDY calculations of AGN broad-lined regions \citep{Korista2004} that were not directly applicable to the TDE conditions. \cite{Gaskell2014} used new CLOUDY calculations to show that a  line optical depth  effects can lead to self-shielding and a suppression of \Ha\ in a way similar to what we
have described. This result, however, was critiqued by \cite{Strubbe2015} who argue that the \cite{Gaskell2014} neglect the large line broadening which should reduce the line optical depth. Indeed, we find that the \Ha\ radial optical depth  is $\lesssim 1$ in our TDE envelopes, and  
that it is  the high  electron scattering optical depth that is critical to the trapping and destruction of hydrogen line photons.

\section{Implications for Observations and Models of TDEs}
\label{sec:ObservationDiscussion}

In many ways, our synthetic spectra resemble those of observed TDE candidates, such as PS1-10jh, near maximum light.  In particular, the spectrum is largely featureless and blue, with two emission lines of HeII and a weak or absent \Ha\ feature.  In detail, however, some discrepancies are apparent.  Our model continua slopes are somewhat shallower, $L_\lambda \propto \lambda^{-3}$, than the Rayleigh-Jeans tail, $L_\lambda \propto \lambda^{-4}$, seen in PS1-10jh \citep{Gezari2012}. Nevertheless, our slope is consistent with HST UV data for ASASSN-14li \citep{Cenko2016}, as well as the spectral indices measured in TDE1/TDE2 \citep{van-Velzen2011}, in the absence of strong dust extinction. 

Our optical luminosity for a massive envelope ($\Menv = 0.5~M_\odot$) tops out at $\approx 10^{43}~\ergss$, which is a factor of $\approx 2$ less than some observed TDEs.  This may in part be due to our incomplete inclusion of metals, which may result in an underestimate of the net
reprocessing efficiency.  The efficiency could also be enhanced if the gas is confined to e.g., a  shell or disk, which would  increase the density. The radiation from an aspherical envelope will also be anisotropic, with the observed luminosity brighter  from viewing angles that maximize the projected surface area.

Observations of ASASSN-14li showed simultaneous x-ray, UV, and optical emission 
\citep{Holoien2016-1,Miller2015-2,Cenko2016}.  The SED cannot be fit by a single blackbody, but appears consistent with the superposition of different temperature blackbodies for the x-rays and optical/UV.  Such an SED resembles our highly ionized
reprocessing envelopes (e.g., Figure~\ref{fig:SEDVaryInnerRadius}) for which 
thermal x-rays characteristic of the hot inner accretion regions are only partially reprocessed to optical radiation characteristic of  the lower envelope temperatures, 
resulting in a multi-blackbody SED.

Our calculations have focused on radiation transport in generic TDE envelopes, remaining largely 
agnostic about the origin of the envelope.  The mechanism for forming an extended gas distribution is unclear, 
and several classes of explanations have been suggested.  
The insights from our radiative transfer studies can be applied (within limit) to illuminate
the possible observable properties of various scenarios.

\subsection{Quasi-static Envelopes}

\citet{Loeb1997}  predicted that TDEs could form an optically thick, roughly spherical, radiation pressure supported envelope with radius $\rout \sim 10^{15}$~cm.
The stability of such an envelope depends on the luminosity being regulated
to be near $L_{\rm Edd}$ \citep{Ulmer1998}.
Analytic studies by \citet{Coughlin2014-1} suggest that even if the 
accretion rate is super-Eddington, the flow can equilibrate in a quasi-stable configuration by allowing energy to escape in a narrowly confined jet. Additionally, the earliest three-dimensional hydrodynamic simulations of TDEs that followed the event past the initial disruption and into the accretion phase, such as \citet{Ayal2000} and \citet{Bogdanovic2004}, provided some evidence for the build-up of material at large radii that could reprocess accretion luminosity. More recent calculations such as \citet{Guillochon2014} \citep[see also][]{Ramirez-Ruiz2009,Rosswog2009} show the development of an extended gas distribution around the eventual accretion disk, which is produced as shock-heated material is lifted above and below the original orbital plane of the disrupted star's orbit; (such calculations, however, have not followed the longer term radiative evolution of this debris).

Our model calculations provide a fair representation of the radiative transfer  in such  quasi-static envelopes, and suggest that such a scenario (with a solar composition envelope of mass $\Menv \approx 0.1-0.5~M_\odot$) can likely  reproduce the maximum light spectra of observed optical TDEs like PS1-10jh.  More specific model calculations are warranted; the power law exponent of the envelope density structure may differ from our choice $\rho \propto r^{-2}$ \citep[for a static radiation pressure-supported envelope, $\rho \propto r^{-3}$ as in][]{Loeb1997}
and the velocity dispersion may be due in part to rotational or circulatory motion, not the random motion adopted in our calculations.

The properties of a quasi-static envelope can be expected to vary over time. Initially, over the timescale for the most-bound debris to fall back to the BH ($t_{\rm fb} \approx 25 [M_{\rm BH}/10^7~M_\odot]^{1/2}$~days for the disruption of a solar mass star), the envelope mass may increase.  For highly-ionized envelopes, a larger \Menv\ leads to more efficient reprocessing (see Figure~\ref{fig:SEDVaryMass}).  The initial rise of the optical TDE light curves could therefore reflect the gradual accumulation of mass in an envelope.
The decline of the light curve with roughly constant color could likewise reflect the draining of the envelope. Such an evolution is hinted at by observations of PS1-10jh, which appears to have a photosphere that first grows and later recedes \citep{Strubbe2015}. The weak dependence of the reprocessed luminosity on the source luminosity, when the mass and extent of the envelope are held fixed, suggests that the peak luminosity of TDEs with $M_{\rm bh} \gtrsim 10^6$~\msun\ may be regulated to be near a few times $10^{43}~\ergss$, with a stronger dependence on the envelope mass than the BH mass.

If, after being assembled, the mass of the envelope remains roughly fixed, then at some
time after maximum light the source luminosity should decline to $L < \Lion$ where recombination fronts form. The source luminosity would then be nearly completely absorbed and reprocessed, and the optical luminosity may track the accretion rate, as
perhaps suggested by the observed $t^{-5/3}$ decline in the optical light curves.
Little soft x-ray emission would be expected at these later times unless deviations
from spherical symmetry allows for channels for x-rays to escape, or the envelope
mass is at  some point depleted.

\subsection{Outflows}
A different mechanism for generating an extended gas distribution
is through outflows.  Super-Eddington accretion onto the BH is likely to
unbind some fraction of the infalling debris \citep{Strubbe2009,Lodato2011,Vinko2015,Metzger2015}.
\citet{Miller2015-1}, drawing on work by \citet{Laor2014},  suggest that line-driven winds analogous to those launched from the atmospheres of massive stars might also be responsible
for launching outflows.   For a wind launched near the tidal disruption radius, the expected escape velocities are $\sim 10^5$ km s$^{-1}$, which is much 
greater than that observed in the line widths of observed TDE spectra.  However,
if debris circularizes at  radii much larger than $R_{\rm td}$, or if the winds are
mass loaded, the expansion velocities may be lower.

 \citet{Strubbe2009} argued that accretion may continuously generate
 an optically thick outflow, which would advect the energy dissipated at the accretion disk to larger radii.  That advection  may circumvent some of the complexities of absorption and reprocessing that we have studied here.  Photons trapped in an outflow will be adiabatically degraded,  which provides a robust mechanism for shifting the SED to longer wavelengths.  The photons finally decouple from the flow at a trapping radius set by $\tau \sim c/v$.   This scenario most closely resembles the calculations  presented here only if  the advection velocity is relatively
low ($v \approx 1000~{\rm km~s^{-1}}$), in which case the trapping radius
coincides with the inner boundary used in our calculations ($\tau \approx 300$).  
Based on analysis of the data of PS1-10jh, \citet{Strubbe2015} suggest an outflow that does move at such a low velocity.   However, their suggested outflow mass of 0.02~$M_\odot$ is likely too low to efficiently reprocess the source radiation, as shown in Figure~\ref{fig:SEDVaryMass}.

\citet{Metzger2015} present a modified outflow picture in which  a large fraction of the inflowing tidal debris is promptly unbound, forming a quasi-spherical outflow with expansion velocities  $\approx 10^4~{\rm km~s^{-1}}$.  Radiation from the accretion disk of the BH is
then absorbed and reprocessed by the outflow.  At times near the peak of the optical
TDE light curve ($\sim$ weeks) that
outflow will have reached radii $\rout \approx 10^{15}$~cm and the diffusion time
is of order the expansion time.  Our models provide a fair
representation of such density distributions, although we do not include
a radial velocity gradient.  Nonetheless, the indications are that such
a scenario can likely reproduce the near
maximum light spectra of observed TDEs. 

In contrast to the quasi-static  or steady wind models, the optical depth of a prompt outflow 
necessarily decreases with time, given that $\rout \propto t$.  
At some point, the envelope should become inefficient at reprocessing, at which time
soft x-rays will escape and the optical light curve may no longer track the accretion rate. Since the critical luminosity
(Equation~\ref{eq:Lcrit}) decreases more steeply ($\Lion \propto t^{-2}$) than
the expected accretion luminosity ($L \propto t^{-5/3}$), the condition $L > \Lion$ is expected to occur at
some late time.  \citet{Metzger2015} have discussed this possible ``ionization break-out'' and estimated its properties.

\subsection{Circularization at Large Radius}

Recent numerical work has suggested that the optical emission might arise from dissipation in material in the process of circularizing at large radii. \citet{Shiokawa2015} performed general relativistic (GR) hydrodynamics simulations and found that shocks located at the apoapse of orbits of returning material lead to build-up of material with enough angular momentum to support wide orbits, with a semi-major axis corresponding to that of the most bound debris from the initial disruption, several times $10^{14}$ cm. This is at least an order of magnitude larger than the periapse of the initial stellar orbit, where most earlier work had suggested that circularization should occur. Similar but less pronounced effects were found in the smoothed particle hydrodynamics simulations including leading-order GR effects performed by \citet{Bonnerot2016-1} and \citet{Hayasaki2015}, and the grid-based Newtonian gravity calculation of \citet{Guillochon2014}. Emission from the circularizing gas at these large radii has been suggested to give rise to the observed optical emission \citep{Piran2015}.

If the gas distribution formed from such circularization processes (or perhaps their
associated outflows) extends from $\sim 10^{14} - 10^{15}$~cm, the situation resembles the numerical calculations presented here.  Given that the gas is highly scattering dominated,
a  relatively high optical depth $\taues \gtrsim 40$
 in that material is  required to thermalize the radiation in the circularization region.  Assuming the optical depth
is at least that high, we find that it makes little difference to the optical spectra whether the luminosity is generated at a  circularization radius of $\sim 10^{14}$~cm or nearer the BH at $\sim 10^{13}$~cm (see Figure~\ref{fig:SEDVaryInnerRadius}).   Indeed, much
more energy is expected to be liberated as material accretes onto the BH
than from the circularization process itself, although mechanisms for hiding
the accretion energy have been suggested \citep{Piran2015,Svirski2015}.

Simultaneous x-ray and optical observations for events such as PTF10iya, ASASSN-14li, and ASASSN-15oi \citep{Cenko2012-1,Holoien2016-1,Miller2015-2,Holoien2016-2} provide some hope at distinguishing
the source of the luminosity, since for highly ionized envelopes the x-ray spectrum will be largely preserved as it propagates through the scattering dominated envelope.  The peak of the x-ray flux then reflects the temperature at the depth at which it was produced (see Figure~\ref{fig:SEDVaryInnerRadius}).

\section{Conclusions}
\label{sec:Conclusions}

\subsection{Summary of Key Results}
This paper has presented analytic estimates and detailed radiative transfer calculations that 
clarify how the spectra of TDEs can be generated within an extended envelope. 
We  have focused our attention on  TDEs around the optical light curve peak, with bolometric luminosities in the range $10^{44} - 10^{45}$ \ergss\ (corresponding to the Eddington luminosity of $10^6 - 10^7 M_\odot$ BHs). 
We accounted for non-LTE effects and the high electron scattering optical depth of the reprocessing material, which we show are crucial for understanding both the thermalization of the optical continuum and the emission line ratios.

 We identified two regimes
of reprocessing, depending on the envelope ionization state.  When the envelope is highly ionized ($L > \Lion$), we find that the intrinsic x-ray emission from the accretion disk is only partially absorbed, giving rise to an SED that peaks in the soft x-ray but is accompanied by an enhanced optical emission component ($\gtrsim 10^{43}$ \ergss\ at wavelengths longer than $1000$~\AA). In addition to providing an optical flux at a sufficient brightness to match observations, this can explain TDEs observed simultaneously in the optical and x-ray, such as PTF10iya, ASASSN-14li, and ASASSN-15oi. 

If the bolometric luminosity declines rapidly enough compared to the mass of the envelope, a critical value can be reached $(L < \Lion)$ for which a helium recombination front forms and the soft x-rays are completely absorbed. This second regime resembles TDEs that have been observed at optical wavelengths without a coincident x-ray signal, such as PS1-10jh and ASASSN-14ae.  In this regime, the reprocessing is completely efficient and 
the optical/UV should closely track the accretion luminosity.

In general, the x-ray through optical SED is not well described by a single blackbody, but is a blend of emission from a variety of depths and temperatures. For this reason, one must be cautious if attempting to fit a single blackbody temperature to optical data. One is likely to underestimate the bolometric luminosity in this way, as the SED can peak at shorter wavelengths than would be inferred from the optical data.

The light curve evolution of TDEs will depend on how the source luminosity, envelope mass, and
envelope radius change with time.
When the envelope is highly ionized ($L > \Lion$), the optical luminosity depends more on the
 envelope mass than the source luminosity.   This is because increasing $L$ leads to higher envelope ionization, which reduces 
the efficiency with which x-rays are reprocessed to the optical.   It is thus possible that the rise and fall of TDE light curves
reflects in large part the growth and subsequent depletion of  mass in the reprocessing envelope.  
As the mass of the envelope decreases, so does the reprocessing efficiency and the optical flux, but the shape of the continuum remains mostly unchanged. This points to an explanation for the near-constant color of the optical continuum. 

 We have demonstrated that  the strength of line features in TDEs
 depends on optically thick radiation transport effects that are not
 well captured by photoionization codes like CLOUDY.
 Even in envelopes of solar composition, the \Ha\ line may be highly suppressed relative to HeII lines due to optical depth effects.
By varying the configuration of material in the reprocessing envelope, a variety of helium-to-hydrogen line ratios can be realized in the optical spectrum. In particular, we have explored what happens as we vary the outer radius of the envelope while keeping the total mass fixed, and have shown a transition of \Ha\ from emission to shallow absorption. Although the radial extent of the envelope is a key parameter, other parameters also influence the line ratios, including the bolometric luminosity and the mass of the envelope.

\subsection{Outstanding Issues}

While our studies have outlined many of the key physical processes at play in 
 TDE envelopes, several questions remain to be addressed.
An obvious area for improvement is relaxing the assumption of spherical symmetry.  Variation in the density and temperature structure of the envelope with
polar angle could lead to important viewing-angle effects which may be important for explaining why soft x-rays are visible in only a subset of observed TDEs. The density and temperature profile of the envelope may also vary with viewing angle, which may have implications for observables such as the slope of the optical continuum.

Another critical area needing further study is the kinematic structure of the envelope.  We have only crudely accounted for  motions via a  Doppler line-width set to a value of order
the virial velocity.  In many scenarios,  velocity gradients are due instead to bulk motions -- outflows or rotation -- that may alter the formation of line features.  In homologously expanding atmospheres, for example, line interactions occur within localized resonance regions, which results in
completely different line optical depth and source function. Radiative
transfer studies of the detailed line profiles may illuminate the kinematics of the
envelope.  Radially expanding outflows like supernovae, for example, generally produce P-Cygni type absorption features, whereas  the optical  lines observed
in TDEs are usually purely in emission, and occasionally  show substructure \citep{Arcavi2014, Holoien2016-1}.

Our calculations have only directly tracked bound-bound and bound-free opacities  from hydrogen, helium, and oxygen. In reality, other metals likely increase the opacity, especially in the x-ray and UV, and ionization edges and lines from these other metals are likely to appear in the spectrum. Nevertheless, for the luminosities we studied, we found that the optical continuum and optical helium-to-hydrogen line ratios do not seem to be greatly influenced by the presence of oxygen, the most abundant metal.

Our assumed inner boundary condition of blackbody radiation emitted at \rin\
deserves further study.  In reality, the source emission spectrum may be that from an accretion disk smaller than \rin.   On the other hand, dense hot gas just below our inner boundary may have a Compton $y$-parameter $\gg 1$, in which case Comptonization  may indeed thermalize the radiation to the  gas temperature near the envelope base. Outside of our inner boundary, the Compton $y$-parameter can be of order unity, and  may have some effect on the spectrum.  The physics of Compton scattering will be directly included
in our future Monte Carlo transport calculations.

Future work will include a more detailed exploration of the parameters governing the spectrum, including the mass of the envelope, accretion disk luminosity, and gas density gradient, as well as how these parameters evolve over time in different scenarios.  Studies of this
sort, in comparison to improved observations of TDEs, will hopefully clarify the
physics governing these transients.

\section*{Acknowledgments}
We thank Janos Botyanszki for code for photoionization cross-sections,
and Tamara Bogdanovi{\'c}, Eric Coughlin, Brad Cenko, Ryan Chornock, Moshe Elitzur, Aleksey Generosov, Julian Krolik, Brian Metzger, Eliot Quataert, Todd Thompson, and Sjoert van Velzen for helpful comments and conversations.  This research used resources of the National Energy
Research Scientific Computing Center, which is supported
by the Office of Science of the U.S. Department
of Energy under Contract No. DE-AC02-05CH11231. DK is supported in part by a Department of Energy Office of Nuclear
Physics Early Career Award, and by the Director, Office of Energy
Research, Office of High Energy and Nuclear Physics, Divisions of
Nuclear Physics, of the U.S. Department of Energy under Contract No.
DE-AC02-05CH11231. 
This work was supported by Einstein grant PF3-
140108 (J. G.), the Packard grant (E. R.), and NASA ATP
grant NNX14AH37G (E. R.)

\appendix

\section{Numerical Method}
\label{sec:Method}

To generate model  spectra of TDE events, we carry out NLTE radiative transport calculations  using the Monte Carlo radiative transfer code SEDONA \citep{Kasen2006,Roth2015-1}.  The calculation is divided into two steps.  First, the multi-wavelength radiation transport is calculated using Monte Carlo method which includes scattering, bound-bound, bound-free, and free-free radiative processes.  Second, the gas temperature and atomic level populations are determined via a solution of the equations of statistical and thermal equilibrium.  Since the photon opacities and emissivities depend on the level populations and temperature, these two steps are iterated (typically 20-60 times) until the envelope structure and output spectra have converged.

For the transport problem, we assume a stationary envelope, which should be applicable for TDE light curves near peak, when the diffusion time through the envelope is less than or comparable to the peak time.  We enforce radiative equilibrium, justified by the heating-time arguments in Section~\ref{sec:analytics}, by ``effectively scattering'' photon packets during each interaction, thereby ensuring energy conservation. At each interaction,  the outgoing packet is reassigned a wavelength, sampled from the NLTE emissivity distribution across all wavelengths, as in \citet{Carciofi2006}.

To calculate the NLTE level populations, we use Monte Carlo estimators of the photoionization rates and the bound-bound radiative rates. We assume statistical equilibrium and solve the set of coupled linear equations such that the net transition rate for each electron level is zero.For hydrogen, we include electron energy levels with principal quantum number of 6 or less. For HeII we include levels with principal quantum number of 9 or less. For oxygen, we use a total of 484 levels across all ionization states. We assume statistical equilibrium and solve the set of coupled linear equations such that the net transition rate for each electron level is zero.

\subsection{Setup and initial conditions}
\label{sec:NumericalDetails}
We divide our spherical grid into 512 zones, equally spaced in radius from zero to \rout. In order to avoid an artificially abrupt truncation of the envelope, we add roughly 100 zones that follow an $r^{-10}$ density profile beyond \rout (we find that our results are not highly sensitive to the exact value of this power-law). We initialized each zone with approximately 1000 photon packets sampled from a blackbody wavelength distribution at the temperature in the zone computed from Equation~\eqref{eq:temp_struct}. During each iteration, we emit approximately 100 million packets at the zone corresponding to our inner radius \rin. 

We impose an absorbing boundary condition at the inner radius --- photons that scatter back below that radius are removed from the calculation. Likewise, photons that escape at the outer radius are tallied and removed from the calculation.

\subsection{Radiative processes included}
\label{sec:RadiativeProcesses}

The radiative processes included in our calculation are electron scattering, free-free (bremsstrahlung), bound-free (photoionization) and bound-bound (line) transitions. 
The extinction coefficient for electron  scattering is $\alpha_{\rm es} = n_e \sigma_t$, while the free-free extinction
coefficient is given by 
\begin{equation}
  \label{eq:FreeFreeAbsorption}
  \alpha^{ff}_\nu = \frac{4 e^6}{3 m h c} \left( \frac{2 \pi}{3 m k T}\right)^{1/2} Z^2 n_e n_{\rm ion} \nu^{-3}\left(1 - e^{-h \nu / k T} \right) .
\end{equation}
By Kirchhoff's law, the free-free emissivity is
$j^{\rm ff}_\nu = \alpha^{\rm ff}_\nu B_\nu(T)$.

The bound-bound extinction coefficient is given by Equation~\eqref{eq:BoundBound}.  The corresponding 
emissivity is
\begin{equation}
  \label{eq:BoundBoundEmissivity}
j^{bb}_{\nu} = \frac{h \nu}{4\pi}n_2 A_{ul} \phi(\nu)
\end{equation}
where $A_{ul}$ is the Einstein coefficient for spontaneous emission, and $\phi(\nu)$ is the line profile. For $\phi(\nu)$, we assume a Gaussian profile with a width corresponding to Doppler velocity of $10^4$ km s$^{-1}$.  

We include bound-free transitions from all excited atomic levels. For hydrogen and HeII, we use the photoionization cross-section \citep{Rybicki1986}
\begin{equation}
  \label{eq:PhotionCSGeneral}
\sigma^{{\rm ion}}_{\nu} = \frac{n_{q,i} \sigma_{0,{\rm H}}}{Z^2} \left(\frac{h\nu}{\chi_{i}}\right)^{-3}    
\end{equation}
where $n_{q,i}$ is the principal quantum number of the bound electron level labeled by index $i$, $\sigma_{0,H} = 6.3 \times 10^{-18}$ cm$^2$ and $\chi_{i}$ is the ionization potential to remove an electron from level $i$.
For the ground state HeI,  we use the photoionization cross-section fits of  \citet{Verner1996}. For oxygen, we use the TOPbase photoionization cross-sections smoothed over resonances. For atomic levels that do not have data, we use an approximate hydrogenic cross-section of the form Equation~\eqref{eq:PhotionCSGeneral}, with $Z$ corresponding to the net nuclear charge seen by the valence electrons  and $n_{q,i}$ the principal quantum number of the valence electron being ionized.

When computing the photoionization extinction coefficient, we include the non-LTE correction for stimulated radiative recombination, yielding \citep[e.g.][]{Mihalas1978-1} 
\begin{eqnarray}
  \label{eq:PhotoionExtinction}
  \alpha_{\nu}^{\rm ion} &=& n_i  \sigma_{\nu}^{\rm ion} \left[1 - \frac{n_e n^+}{n_i} f(T)\exp\left(-\frac{h \nu}{k_B T} \right) \right] \, \, ,\nonumber \\
  f(T) &=& \left(\frac{h^2}{2 \pi m_e k_B T}\right)^{3/2} \frac{g^-}{2 g^+}\exp \left(\frac{\chi_{i}}{k_B T}\right)
\end{eqnarray}
where $n_i$ is the number density of particles with bound electron in level $i$, $n^+$ is the number density for the ions in the ground state of the next highest ionization state, $g^-$ and $g^+$ are the statistical weights of the species being ionized and the ionized state, respectively, and $T$ is the temperature of the free electrons. 

This opacity must be summed over all elements and all bound electron levels labeled by the index $i$ within each ionization state.

To derive radiative recombination cross-sections, $\sigma^{\rm rec}(u_e)$, as a function of electron speed $u_e$, we use the Milne relations which relate $\sigma^{\rm rec}(u_e)$ to the photoionization cross-sections. The associated emissivity for bound-free recombination is \citep[see][]{Osterbrock2006}.
\begin{eqnarray}
  \label{eq:RecombinationEmissivity}
  j^{\rm recomb}_{\nu} &=& \frac{n^+ n_e}{4 \pi} u_e f_u\sigma^{{\rm rec}}(u_e) h \nu \frac{du}{d\nu}\, \, , \nonumber \\
u_e &=& \sqrt{\frac{2}{m_e} (h \nu - \chi_{i})} \, \, , \nonumber \\
f_u &=& \frac{4}{\sqrt{\pi}} \left(\frac{m_e}{2 k T}\right)^{3/2} u_e^2 \exp \left(- \frac{m_e u_e^2}{ 2 k T}\right).
\end{eqnarray}
 This emissivity must be summed over all elements and all bound electron levels to which the free electron may recombine. 
We derive the temperature-dependent radiative recombination coefficient for each atomic level by integrating $\sigma^{\rm rec}(u_e)$ over
the electron Maxwell-Boltzmann distribution.

\end{document}